\documentclass[10pt,aps,pra,twocolumn,superscriptaddress,floatfix,nofootinbib]{revtex4-2}

\usepackage{amsmath,amssymb,bm}
\usepackage[T1]{fontenc}
\usepackage{bookman}
\usepackage{graphicx}
\usepackage{xcolor}
\definecolor{linkcol}{RGB}{20,40,160}
\definecolor{citecol}{RGB}{20,90,170}
\definecolor{urlcol}{RGB}{0,110,135}
\usepackage[colorlinks=true,linkcolor=linkcol,citecolor=citecol,urlcolor=urlcol]{hyperref}
\usepackage{orcidlink}

\newcommand{\Wb}{\mathcal{W}}
\newcommand{\Cc}{\mathcal{C}}
\newcommand{\bra}[1]{\langle #1|}
\newcommand{\ket}[1]{|#1\rangle}
\newcommand{\ip}[2]{\langle #1|#2\rangle}

\begin{document}

\title{\texorpdfstring{Power-resolved ergotropy and correlation redistribution
in a bosonic quantum battery driven by a coherent-state superposition}{Power-resolved ergotropy and correlation redistribution in a bosonic quantum battery driven by a coherent-state superposition}}

\author{João Pedro d'El-Rey
\href{https://orcid.org/0009-0009-2046-6346}{\includegraphics[scale=0.04]{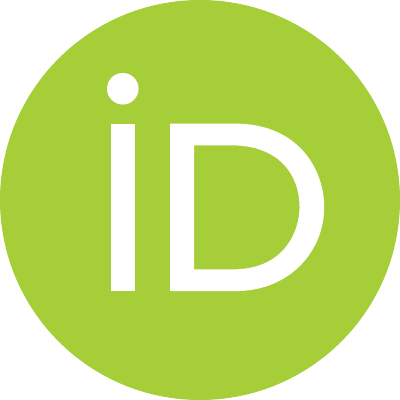}}
}
\affiliation{Brazilian Center for Physics Research, Rio de Janeiro, RJ, 22290-180, Brazil}
\email{jpbhdrey@cbpf.br}

\author{Tailan S. Sarubi
\href{https://orcid.org/0009-0009-2046-6346}
{\includegraphics[scale=0.04]{orcidid.pdf}}
}
\affiliation{Physics Department, Federal University of Rio Grande do Norte,
Natal, Rio Grande do Norte, 59072-970, Brazil}
\affiliation{International Institute of Physics,
Federal University of Rio Grande do Norte,
Natal, Rio Grande do Norte, 59072-970, Brazil}

\author{Ana~C.~S.~Costa\,\orcidlink{0000-0002-4014-0695}}
\affiliation{Department of Physics, Federal University of Paran\'a, Curitiba, Paran\'a, P.O. Box 19044, 81531-980, Brazil}

\author{José~G.~G.~de Oliveira,~Jr.\href{https://orcid.org/0000-0002-1490-8859}{\includegraphics[scale=0.04]{orcidid.pdf}}}
\affiliation{NEMeS, Department of Exact Sciences, State University of Santa Cruz, Ilh\'eus, Bahia, 45662-900, Brazil}
\affiliation{Department of Physics, Federal University of Paran\'a, Curitiba, Paran\'a, P.O. Box 19044, 81531-980, Brazil}

\date{\today}

\begin{abstract}

Ergotropic-gap relations provide thermodynamic witnesses of quantum correlations, but a direct dynamical connection between extractable work and entanglement remains largely unexplored in multipartite non-Gaussian bosonic systems. Here, we establish such a connection for a bosonic quantum battery driven by a coherent-state superposition. Despite the infinite-dimensional Hilbert space, the dynamics retains a simple structure that allows the relevant work and correlation measures to be obtained exactly. We show that the loss of extractable work associated with the passive part of the battery energy is directly determined by charger-battery entanglement. This relation also separates the charging power into contributions from energy transfer and from changes in the battery spectrum. At the same time, multipartite correlations are progressively redistributed from charger-involving correlations to correlations internal to the battery. For a symmetric resonant protocol, complete energy transfer coincides with charger-battery disentanglement and fully extractable stored energy, while multipartite correlations within the battery are maximal. Individual cells, however, remain mixed, revealing a distinction between global and local work extraction. We further show that, in the regime considered, locally extractable work is entirely supported by energetic coherence. Under single-photon loss, the analytical structure survives, but environmental correlations separate charger disentanglement from maximal work extraction.
\end{abstract}

\maketitle

\section{Introduction}

Quantum batteries store and release energy under controlled unitary operations~\cite{Alicki2013,Allahverdyan2004,Hovhannisyan2013,Binder2015,Campaioli2017,Ferraro2018,Andolina2018,Andolina2019,Rossini2019,Campaioli2024colloquium,Ferraro2026review}; however, stored energy alone does not characterize their performance. The relevant thermodynamic quantity is the ergotropy, the maximum work extractable by cyclic unitary operations~\cite{Allahverdyan2004,Pusz1978,Lenard1978}. Because unitaries preserve the spectrum of the density operator, the ergotropy equals the difference between the actual energy of the state and the energy of its passive rearrangement, in which populations are reordered against the energy spectrum~\cite{Allahverdyan2004}.

Quantum correlations act on this quantity in two distinct ways. Entanglement and collective operations can raise charging power and produce advantages in charging time~\cite{Binder2015,Campaioli2017,Ferraro2018,Gyhm2022}. Correlations between a battery and an external party instead render the reduced battery state mixed and can lock part of the stored energy away from local extraction~\cite{Hovhannisyan2013,Andolina2019,Rossini2019,Salvia2023,Castellano2024,Francica2022qcoh}, although correlated initial states can restore lossless transport in dedicated settings~\cite{Simon2025lossless}. The tension is particularly direct in charger-mediated batteries~\cite{Andolina2018,Farina2019,Pushpan2025}, where energy transfer and entanglement generation occur simultaneously and the latter reshapes the passive spectrum of the reduced battery state.

The ergotropic gap, the difference between work extractable by global and by local unitaries, quantifies this trade-off and has become a useful witness of correlations~\cite{Mukherjee2016,Alimuddin2019,Alimuddin2020passive}. For three-qubit pure states, an exact relation connects bipartite ergotropic gaps and concurrence~\cite{Puliyil2022}, with multipartite extensions and entanglement-class diagnostics following~\cite{Yang2024polytope,Yang2024bipartite}. For continuous-variable (CV) systems, a relative ergotropic gap certifies entanglement in Gaussian states and, through the Shchukin--Vogel criterion, in photon-subtracted non-Gaussian states~\cite{PoloRodriguez2026}, while multimode Gaussian constructions build hierarchies of multipartite entanglement from related thermodynamic quantities~\cite{SamantaHierarchies2026}.

An important regime remains outside this framework: a dynamical, multipartite, non-Gaussian CV battery whose reduced spectra are not determined by a covariance matrix. Fock-space truncation is then the usual analytical bottleneck. Existing bosonic models often remain Gaussian~\cite{Konar2024multimode,Cavaliere2025,Mpemba2025gaussian}, restrict to bipartite charger-battery settings~\cite{Andolina2018,Farina2019,Pushpan2025}, exploit anharmonic couplings~\cite{Andolina2025anharmonic}, or treat multipartite structure numerically~\cite{Beder2025cavityarray}.

We consider a harmonic-oscillator charger, initially prepared in a coherent-state superposition, coupled through a number-conserving star interaction to three oscillator cells in vacuum. Superpositions of distinct, nonorthogonal coherent-state branches and their effective finite-dimensional descriptions are well established in the entangled-coherent-state literature~\cite{Sanders1992,Sanders2012}. The novelty sought here is therefore not the two-branch encoding itself, but the exact thermodynamic and correlation identities that follow when this structure is combined with charger-mediated battery dynamics. The quadratic evolution preserves a two-branch superposition of product coherent states. Consequently, every reduced state used below has support on at most two branch vectors, irrespective of the number of populated Fock levels. Exact orthonormalization of these supports maps every bipartition relevant for concurrence to an effective two-qubit state without truncating the physical oscillator Hilbert spaces.

This structure yields a closed chain from two-branch dynamics to rank-two spectra, exact Wootters concurrences, and work-extraction identities. First, all bipartite concurrences appearing in the work are Wootters concurrences on exact $2\times2$ supports. Their closed overlap formula gives two squared-concurrence monogamy residuals in the sense of Coffman--Kundu--Wootters and its multiqubit extension~\cite{CKW2000,Osborne2006}: a battery-internal residual $\mathcal R_i^{(B)}$ and a charger-involving residual $\mathcal R_i^{(C)}$. The first is, in general, a mixed-state monogamy residual rather than the convex-roof three-tangle; it becomes the standard three-tangle when the battery is pure. During unitary evolution, the second is exactly the three-tangle of the effective pure tripartition $B_i|C|B_jB_k$. Second, charger-battery concurrence determines the passive spectral penalty exactly in the unitary problem, and the corresponding power identity separates excitation flow from spectral purification. Third, in the symmetric resonant even-cat protocol, we prove that both global and single-cell ergotropies increase monotonically with the transferred fraction on the charging branch, while the single-cell Fock diagonal is passive throughout the parameter regime used in the figures.

Single-photon loss leaves the two-branch rank-two support intact but changes what the passive penalty represents. Introducing the environment-overlap factor $x_E$, we obtain the exact mixedness decomposition
\begin{equation}
 \mathcal M_B^2=\mathcal C_{C:B}^2+\mathcal A(1-x_E)(1-x_B),
 \label{eq:intro-mixedness}
\end{equation}
Equation~\eqref{eq:intro-mixedness} separates charger-battery entanglement from the mixedness generated by which-branch information leaked to the environment. Under uniform loss, battery-only spectral quantities retain the lossless functional dependence after $\Theta\to\widetilde\Theta$. Charger-involving correlations, by contrast, depend separately on the system and environment overlaps. This distinction splits the lossless coincidence between charger disentanglement and maximal extraction into two events.

The coherent--incoherent decomposition further sharpens the global--local distinction. For the representative even-cat protocol with $\alpha_0=1.5$, each single-cell diagonal state is passive, so all locally extractable work is coherence-supported, whereas the full battery develops a finite incoherent contribution accessible only through collective operations. Section~\ref{sec:model} introduces the model, Sec.~\ref{sec:rank2} develops the rank-two spectral and work-extraction structure, Sec.~\ref{sec:penalty} establishes the concurrence and monogamy relations, Sec.~\ref{sec:symmetric} specializes to the symmetric resonant even-cat protocol, Sec.~\ref{sec:diss} treats single-photon loss, and Sec.~\ref{sec:disc} summarizes the implications and limitations.

\section{Model}\label{sec:model}

\begin{figure}[t]
 \centering
 \includegraphics[width=\columnwidth]{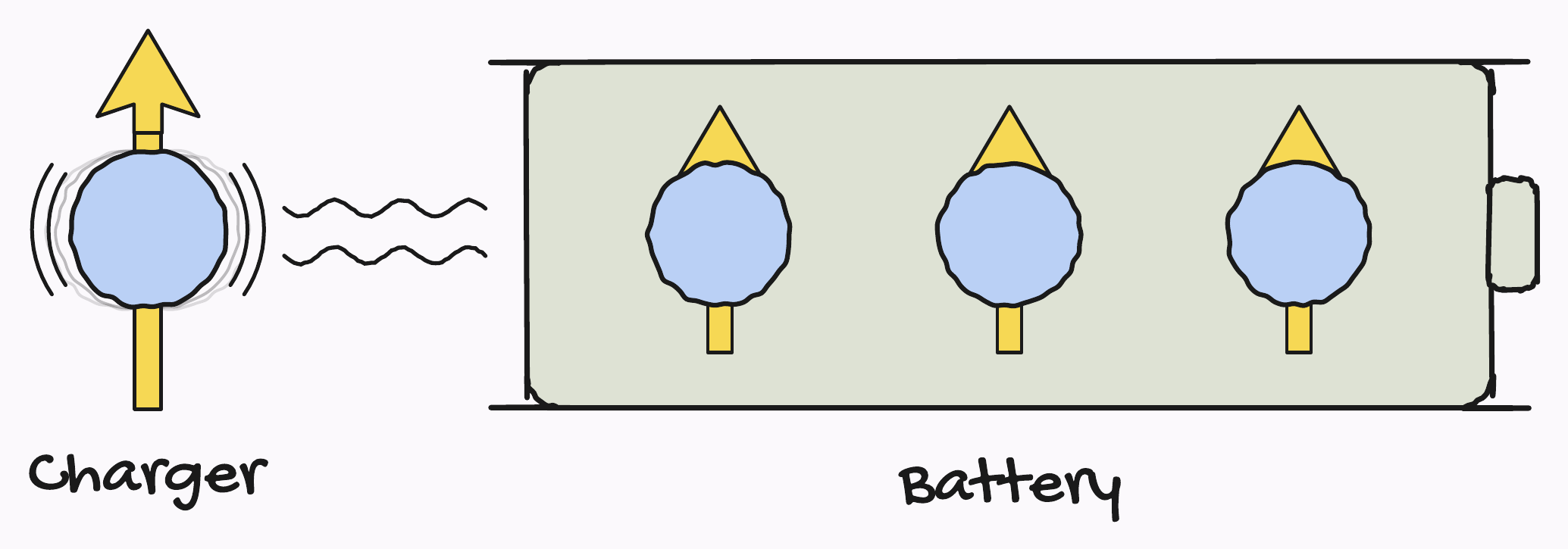}
 \caption{Schematic illustration of the charger-mediated bosonic battery model. A single harmonic-oscillator charger $C$ is coupled to three harmonic-oscillator battery cells; the three battery oscillators represent $B_1$, $B_2$, and $B_3$ from left to right. Excitations are redistributed through the number-conserving interaction of Eq.~\eqref{eq:H}.}
 \label{fig:schematic}
\end{figure}

We consider a charger-mediated bosonic battery composed of one harmonic oscillator $C$ coupled to three oscillator cells $B_1,B_2,B_3$, as illustrated in Fig.~\ref{fig:schematic}. The Hamiltonian is a charger-mediated specialization~\cite{Andolina2018,Farina2019} of the multi-oscillator coherent-state model whose entanglement dynamics was solved analytically for initial coherent states in earlier work~\cite{DePaula2014}:
\begin{equation}
  \hat H=\hbar\omega_c\hat c^\dagger\hat c+\hbar\sum_{i=1}^3\omega_i\hat b_i^\dagger\hat b_i+\hbar\sum_{i=1}^3\gamma_i(\hat c^\dagger\hat b_i+\hat c\,\hat b_i^\dagger),
  \label{eq:H}
\end{equation}
where $\hat c$ ($\hat c^\dagger$) is the charger annihilation (creation) operator and $\hat b_i$ ($\hat b_i^\dagger$) is the corresponding operator for cell $B_i$. Here $\omega_c$ is the charger frequency and $\omega_i$ the frequency of the $i$th cell.

The Hamiltonian conserves the total excitation number $\hat N=\hat c^\dagger\hat c+\sum_i\hat b_i^\dagger\hat b_i$, so the interaction only redistributes excitations between charger and cells. For work extraction we use the free battery Hamiltonian
\begin{equation}
  \hat H_B=E_0\mathbb I_B+\widetilde{\hat H}_B
  =\sum_{i=1}^3\hbar\omega_i\Bigl(\hat b_i^\dagger\hat b_i+\tfrac12\Bigr),
  \label{eq:HB}
\end{equation}
where $E_0=\tfrac{\hbar}{2}\sum_i\omega_i$ and $\widetilde{\hat H}_B=\hbar\sum_i\omega_i\hat b_i^\dagger\hat b_i$. Since $\operatorname{Tr}\rho_B=1$, the stored energy above the zero-point contribution is
\begin{equation}
 \Delta E_B(t)=\operatorname{Tr}[\hat H_B\rho_B(t)]-E_0
 =\operatorname{Tr}[\widetilde{\hat H}_B\rho_B(t)].
\end{equation}

The charger starts in a coherent-state superposition and the cells in vacuum,
\begin{equation}
  \ket{\Psi(0)}=\mathcal N_0\bigl(c_\alpha\ket{\alpha_0}_C+c_\beta\ket{\beta_0}_C\bigr)\ket{0,0,0}_B,
  \label{eq:init}
\end{equation}
where we impose the coefficient convention $|c_\alpha|^2+|c_\beta|^2=1$. This convention does not normalize a superposition of nonorthogonal coherent states; the physical normalization is carried by $\mathcal N_0$, chosen real and positive without loss of generality, with
\begin{equation}
 |\mathcal N_0|^{-2}
 =1+c_\alpha^*c_\beta\langle\alpha_0|\beta_0\rangle
   +c_\alpha c_\beta^*\langle\beta_0|\alpha_0\rangle.
 \label{eq:N0}
\end{equation}
Because $\hat H$ is quadratic and number conserving, each coherent product remains a coherent product, and the state retains exactly two branches,
\begin{equation}
  \ket{\Psi(t)}=\mathcal N_0\bigl(c_\alpha\ket{\alpha_t}_C\ket{\lambda_t}_B+c_\beta\ket{\beta_t}_C\ket{\chi_t}_B\bigr),
  \label{eq:psit}
\end{equation}
where $\ket{\lambda_t}_B=\bigotimes_i\ket{\lambda_t^{(i)}}$ and $\ket{\chi_t}_B=\bigotimes_i\ket{\chi_t^{(i)}}$. The coherent amplitudes obey the single-particle equation $i\dot{\bm z}=M\bm z$ and hence $\bm z(t)=e^{-iMt}\bm z(0)$, with
\begin{equation}
  M=
  \begin{pmatrix}\omega_c&\gamma_1&\gamma_2&\gamma_3\\ \gamma_1&\omega_1&0&0\\ \gamma_2&0&\omega_2&0\\ \gamma_3&0&0&\omega_3
  \end{pmatrix}.
  \label{eq:M}
\end{equation}
The two branch vectors start from $(\alpha_0,0,0,0)^T$ and $(\beta_0,0,0,0)^T$. Since $M=M^\dagger$, $e^{-iMt}$ is unitary. Consequently, the full multimode branch overlap is conserved:
\begin{equation}
 \langle\bm z^\alpha(t)|\bm z^\beta(t)\rangle
 =\langle\bm z^\alpha(0)|\bm z^\beta(0)\rangle
 =\langle\alpha_0|\beta_0\rangle.
 \label{eq:overlapconservation}
\end{equation}
Writing this overlap as the product of charger and battery overlaps gives
\begin{equation}
 s(t)q_B(t)=\langle\alpha_0|\beta_0\rangle.
 \label{eq:sqconst}
\end{equation}
Thus $s(t)$ and $q_B(t)$ generally vary separately as branch distinguishability is transferred, but their product and therefore the normalization $\mathcal N_0$ are time independent throughout the unitary dynamics. The propagation is summarized in Appendix~\ref{app:prop}.

We often specialize to the symmetric resonant regime $\omega_c=\omega_1=\omega_2=\omega_3\equiv\omega$ and $\gamma_1=\gamma_2=\gamma_3\equiv\gamma$. In the interaction picture,
\begin{equation}
 \alpha_t=\alpha_0e^{-i\omega t}f(t),\qquad
 \lambda_t^{(i)}=\alpha_0e^{-i\omega t}g(t),
\end{equation}
with analogous relations for the $\beta$ branch and
\begin{equation}
  f(t)=\cos(\sqrt3\,\gamma t),\qquad g(t)=-\tfrac{i}{\sqrt3}\sin(\sqrt3\,\gamma t).
  \label{eq:fg}
\end{equation}
The factor $\sqrt3$ is the collective transfer frequency of the three-cell star. For $N$ identical resonant cells, the charger couples only to the bright combination $\hat{\bar b}=\sum_i\hat b_i/\sqrt N$ with strength $\sqrt N\gamma$, while the $N-1$ dark combinations remain unpopulated for vacuum initial conditions. Then $f(t)=\cos(\sqrt N\gamma t)$ and $g(t)=-(i/\sqrt N)\sin(\sqrt N\gamma t)$. We keep $N=3$ because it is the smallest case supporting the tripartite monogamy structure studied below.

The charger and battery excitation weights are $\Xi(t)=|f|^2=\cos^2(\sqrt3\gamma t)$ and $\Theta(t)=3|g|^2=\sin^2(\sqrt3\gamma t)$, with $\Xi+\Theta=1$. The first full-transfer time is
\begin{equation}
 t_*=\frac{\pi}{2\sqrt3\gamma},
\end{equation}
where $\Theta(t_*)=1$ and $f(t_*)=0$.

The thermodynamic figure of merit for the battery is not the stored energy alone but the fraction extractable by cyclic unitary operations. For a state $\rho$ and Hamiltonian $\hat H$, the ergotropy is~\cite{Allahverdyan2004}
\begin{equation}
  \Wb(\rho,\hat H)=\operatorname{Tr}(\hat H\rho)-\min_U\operatorname{Tr}\bigl(\hat H U\rho U^\dagger\bigr),
  \label{eq:ergo-def}
\end{equation}
where the minimum is over all unitaries. It is attained by a passive state $\rho^{(p)}$~\cite{Pusz1978,Lenard1978}. Since unitary operations preserve the spectrum of $\rho$, passivation is achieved by placing the largest eigenvalues on the lowest available energies~\cite{Allahverdyan2004}. Allowing arbitrary unitaries on $B_1B_2B_3$ defines the global ergotropy, whereas restricting extraction to independent operations on each cell gives the corresponding local ergotropies. The difference between these operational settings is where correlations internal to the battery become thermodynamically relevant.

\section{Rank-two structure and work extraction}\label{sec:rank2}

Every reduced state used for ergotropy or concurrence is supported by at most two coherent-state branches, despite the infinite-dimensional oscillator Hilbert spaces. Define the branch overlaps
\begin{equation}
  s\equiv\ip{\alpha_t}{\beta_t},\quad q_i\equiv\ip{\lambda^{(i)}_t}{\chi^{(i)}_t},\quad q_B\equiv\ip{\lambda_t}{\chi_t}=q_1q_2q_3,
  \label{eq:overlaps}
\end{equation}
The exact coherent-state overlap is
\begin{equation}
 \ip{\mu}{\nu}=\exp\!\left(-\tfrac12|\mu|^2-\tfrac12|\nu|^2+\mu^*\nu\right),
 \label{eq:coherent-overlap}
\end{equation}
so $|\ip{\mu}{\nu}|^2=e^{-|\mu-\nu|^2}$. The phases of $s$ and $q_i$ enter the interference terms and are retained throughout. Equation~\eqref{eq:sqconst} implies that
\[
 |\mathcal N_0|^{-2}=1+c_\alpha^*c_\beta s q_B+c_\alpha c_\beta^*s^*q_B^*,
\]
which is identical to Eq.~\eqref{eq:N0} at every time.

Tracing out the charger from the global state~\eqref{eq:psit} gives
\begin{align}
  \rho_B=|\mathcal N_0|^2\bigl[\,&|c_\alpha|^2\ket{\lambda_t}\bra{\lambda_t}+|c_\beta|^2\ket{\chi_t}\bra{\chi_t}\nonumber\\
  &+c_\alpha c_\beta^* s^*\ket{\lambda_t}\bra{\chi_t}+c_\alpha^* c_\beta s\ket{\chi_t}\bra{\lambda_t}\bigr],
  \label{eq:rhoB}
\end{align}
so $\operatorname{supp}\rho_B\subseteq\operatorname{span}\{\ket{\lambda_t},\ket{\chi_t}\}$ and $\operatorname{rank}\rho_B\le2$. Its determinant on this support is
\begin{equation}
  \det\rho_B=|\mathcal N_0|^4|c_\alpha|^2|c_\beta|^2(1-|s|^2)(1-|q_B|^2),
  \label{eq:detB}
\end{equation}
and, with $\mathrm{Tr}\,\rho_B=1$, the only nonzero eigenvalues are
\begin{equation}
  \nu_\pm=\tfrac12\Bigl[1\pm\sqrt{1-4\det\rho_B}\Bigr].
  \label{eq:nupm}
\end{equation}
For a single cell, write $Q_i\equiv\prod_{\ell\ne i}q_\ell$ so that $q_B=q_iQ_i$. Tracing out $C$ and the other two cells yields the rank-two state
\begin{align}
  \rho_{B_i}=|\mathcal N_0|^2\bigl[\,&|c_\alpha|^2\ket{\lambda^{(i)}_t}\bra{\lambda^{(i)}_t}
    +|c_\beta|^2\ket{\chi^{(i)}_t}\bra{\chi^{(i)}_t}\nonumber\\
    &+c_\alpha c_\beta^* s^* Q_i^*\ket{\lambda^{(i)}_t}\bra{\chi^{(i)}_t}\nonumber\\
  &+c_\alpha^* c_\beta s Q_i\ket{\chi^{(i)}_t}\bra{\lambda^{(i)}_t}\bigr],
  \label{eq:rhoBi}
\end{align}
with determinant
\begin{equation}
  \det\rho_{B_i}=|\mathcal N_0|^4|c_\alpha|^2|c_\beta|^2(1-|s|^2|Q_i|^2)(1-|q_i|^2)
  \label{eq:detBi}
\end{equation}
and eigenvalues $\nu^{(i)}_\pm=\tfrac12[1\pm\sqrt{1-4\det\rho_{B_i}}]$. Equations~\eqref{eq:nupm} and~\eqref{eq:detBi} are the spectral backbone of the analysis, obtained without truncating any oscillator Hilbert space.

A remark on the nature of this reduction is in order. Because each reduced state lives on the span of two nonorthogonal coherent branches, an exact orthonormalization of that span maps it to an effective qubit operator (Appendix~\ref{app:tangles}), on which determinants, spectra, and concurrences are evaluated in closed form. This reduction is a kinematic property of the dynamically generated state, not of the Hamiltonian: $\hat H$ possesses no two-dimensional invariant subspace, and the effective-qubit bases themselves rotate in time together with the branches. The dynamics is therefore always solved in the full oscillator space, at the level of the coherent amplitudes of Eq.~\eqref{eq:M}; the qubit map enters only afterwards, as an exact bookkeeping device for the spectral and entanglement content of the reduced states.

\subsection{Passive states and global and local ergotropies}\label{sec:passive}

Both the full battery state and the single-cell states have only two nonzero eigenvalues, so their passive rearrangements require only the lower spectral weight: for a rank-two state with $p_+\ge p_-$, the passive reordering of Eq.~\eqref{eq:ergo-def} places $p_+$ on the ground level and $p_-$ on the first excited level.

For the full battery, let $\omega_{\min}=\min\{\omega_1,\omega_2,\omega_3\}$. The first excitation above the vacuum has energy $E_0+\hbar\omega_{\min}$, so the passive state is $\rho^{(p)}_B=\nu_+\ket{0}\bra{0}+\nu_-\ket{\phi_1}\bra{\phi_1}$ with $\ket{\phi_1}$ any normalized vector in the first-excited eigenspace. If that level is degenerate, as in the symmetric case where $\ket{\phi_1}$ may be any superposition of $\ket{1,0,0}$, $\ket{0,1,0}$, $\ket{0,0,1}$, the passive eigenvectors are not unique but the passive energy $E_0+\hbar\omega_{\min}\nu_-$ is. With the stored excitation energy $\Delta E_B=\hbar\,\varepsilon_B$ from Appendix~\ref{app:energy},
\begin{equation}
  \Wb_B=\Delta E_B-\hbar\omega_{\min}\nu_-=\hbar(\varepsilon_B-\omega_{\min}\nu_-).
  \label{eq:WB}
\end{equation}
The smaller eigenvalue of the reduced battery state is therefore the passive energy itself, in units of the lowest gap. The global ergotropic efficiency is 
\begin{equation}
    \eta_B=\Wb_B/\Delta E_B=1-\omega_{\min}\nu_-/\varepsilon_B,
    \label{eq:effi}
\end{equation} 
so unit efficiency is equivalent to $\nu_-=0$ whenever $\Delta E_B>0$.

The same logic applies to each cell, with one operational distinction. The local ergotropy $\Wb_{B_i}$ is the work extractable by unitaries on $B_i$ alone, the single-cell instance of the product-unitary restriction introduced in Sec.~\ref{sec:model}. With $\hat H_{B_i}=\hbar\omega_i(\hat b_i^\dagger\hat b_i+\tfrac12)$ and stored energy $\Delta E_{B_i}=\hbar\omega_i\bar n_i$ (Appendix~\ref{app:energy}),
\begin{equation}
  \Wb_{B_i}=\Delta E_{B_i}-\hbar\omega_i\nu^{(i)}_-=\hbar\omega_i\bigl(\bar n_i-\nu^{(i)}_-\bigr),
  \label{eq:WBi}
\end{equation}
and $\eta_{B_i}=1-\nu^{(i)}_-/\bar n_i$. Equations~\eqref{eq:WB} and~\eqref{eq:WBi} share the same spectral form but answer different operational questions: Eq.~\eqref{eq:WB} allows arbitrary unitaries on $B_1B_2B_3$, Eq.~\eqref{eq:WBi} only a unitary on a single oscillator. The difference between local and global operations lies in the correlations that can leave the full battery active under collective extraction while making the individual cells locally mixed and partly passive.

\subsection{Coherent and incoherent work content}\label{sec:cohinc}

The passive penalties $\nu_-$ and $\nu_-^{(i)}$ quantify how much of the stored energy is locked by mixedness, but they do not reveal which physical resource carries the work that remains extractable. Part of the ergotropy can be harvested by permutations of energy populations; the rest requires coherence between distinct energy sectors and is lost under energetic dephasing. Following the coherent--incoherent ergotropy split of Francica \emph{et al.}~\cite{Francica2020coherent}, we separate these two contributions. In the present multimode problem, however, the battery Hamiltonian can possess degenerate energy levels. A complete dephasing in an arbitrarily chosen basis inside each degenerate sector would make the split basis dependent. We therefore use a degeneracy-respecting extension based on spectral pinching onto complete energy eigenspaces. In the present model, this refinement answers a question the rank-two spectra leave open: whether the work stored in the battery, globally and cell by cell, is population-based or coherence-based.

Let $\hat H_B=\sum_E E\Pi_E$ be the spectral decomposition of the battery Hamiltonian, where $\Pi_E$ projects onto the full eigenspace of energy $E$. We define the degeneracy-respecting energy-dephasing (pinching) channel
\begin{equation}
 \mathcal D_{\hat H_B}(X)=\sum_E\Pi_E X\Pi_E,
 \label{eq:dephasechannel}
\end{equation}
which is completely positive, trace preserving, unital, and idempotent. The dephased battery state is $\rho_B^{\rm diag}=\mathcal D_{\hat H_B}(\rho_B)$. If an energy is degenerate, Eq.~\eqref{eq:dephasechannel} removes coherences only between distinct energy sectors and leaves the operator inside each degenerate eigenspace unchanged. The resulting split is therefore invariant under basis changes within an exactly degenerate sector. Equivalently, one may diagonalize $\rho_B$ inside each degenerate energy block when constructing the passive rearrangement; coherences between exactly degenerate vectors do not contribute to the energetic coherence resource.
Then
\begin{equation}
  \Wb_B=\Wb_B^{\rm coh}+\Wb_B^{\rm inc},\qquad
  \Wb_B^{\rm inc}=\Wb(\rho_B^{\rm diag},\hat H_B),
  \label{eq:cohincglobal}
\end{equation}
with $\Wb_B^{\rm coh}\ge0$ the work that is inaccessible after removing energetic coherence. If $\{\pi_r^\downarrow\}$ are the eigenvalues of the energy blocks $\Pi_E\rho_B\Pi_E$, collected in nonincreasing order, and $\{\epsilon_r^\uparrow\}$ are the excitation energies of $\widetilde{\hat H}_B$ in nondecreasing order, including their degeneracies, then
\begin{equation}
  \Wb_B^{\rm inc}=\Delta E_B-\sum_r \pi_r^\downarrow\epsilon_r^\uparrow,
  \qquad
  \Wb_B^{\rm coh}=\sum_r \pi_r^\downarrow\epsilon_r^\uparrow-\hbar\omega_{\min}\nu_-.
  \label{eq:globalcohinc}
\end{equation}
When the many-body spectrum of $\hat H_B$ is nondegenerate, or after any accidental degeneracies are resolved, these block eigenvalues reduce to the product-Fock populations
\begin{align}
  P_{\bm n}=|\mathcal N_0|^2\Bigl[&|c_\alpha|^2\prod_i p_{\lambda_i}(n_i)+|c_\beta|^2\prod_i p_{\chi_i}(n_i)\nonumber\\
  &+2\Re\!\left(c_\alpha c_\beta^*s^*\prod_iK_i(n_i)\right)\Bigr],
  \label{eq:pops}
\end{align}
where
\begin{equation}
 p_z(n)=e^{-|z|^2}\frac{|z|^{2n}}{n!},\qquad
 K_i(n)=\langle n|\lambda_t^{(i)}\rangle\langle\chi_t^{(i)}|n\rangle.
 \label{eq:local-poisson-kernel}
\end{equation}

In the symmetric resonant even-cat protocol, the excitation-$n$ eigenspace has degeneracy $d_n=\binom{n+2}{2}$. The two projected coherent branches differ only by the parity factor $(-1)^n$, so each block $\Pi_n\rho_B\Pi_n$ has rank one. With $a=|\alpha_0|^2$ and $y=\Theta(t)$, its only nonzero eigenvalue is
\begin{equation}
  \pi_n(t)=\frac{e^{-ay}(ay)^n}{n!\left(1+e^{-2a}\right)}
  \left[1+(-1)^n e^{-2a(1-y)}\right].
  \label{eq:globalblockpops}
\end{equation}
The remaining $d_n-1$ eigenvalues in that energy sector vanish. Equation~\eqref{eq:globalcohinc} is therefore evaluated by sorting the sequence $\{\pi_n\}$ together with these zero eigenvalues against the spectrum $\{\hbar\omega n\}$ with multiplicities $d_n$. This treatment removes the basis ambiguity associated with the symmetric degeneracies. We verified Eq.~\eqref{eq:globalblockpops} against an exact diagonalization of every energy block of $\rho_B$ in a truncated three-mode Fock space, confirming both the rank-one structure and the eigenvalue to machine precision.

The same construction applies to the $i$th cell and completes the global--local comparison. Its dephased state is
\begin{equation}
  \rho_{B_i}^{\rm diag}=\sum_{n=0}^\infty p_n^{(i)}\ket n\bra n,
\end{equation}
with
\begin{align}
  p_n^{(i)}=|\mathcal N_0|^2\Bigl[&|c_\alpha|^2p_{\lambda_i}(n)+|c_\beta|^2p_{\chi_i}(n)\nonumber\\
  &+2\Re\!\left(c_\alpha c_\beta^*s^*Q_i^*K_i(n)\right)\Bigr].
  \label{eq:localpops}
\end{align}
Writing $p_n^{(i)\downarrow}$ for the same populations in nonincreasing order gives
\begin{align}
  \Wb_{B_i}^{\rm inc}&=\hbar\omega_i\left(\bar n_i-\sum_{n=0}^\infty n\,p_n^{(i)\downarrow}\right),\label{eq:WincBi}\\
  \Wb_{B_i}^{\rm coh}&=\hbar\omega_i\left(\sum_{n=0}^\infty n\,p_n^{(i)\downarrow}-\nu_-^{(i)}\right).
  \label{eq:WcohBi}
\end{align}
Thus the local passive penalty $\hbar\omega_i\nu_-^{(i)}$ and the local coherence contribution answer different questions: the first quantifies mixedness relative to the rest of the system, whereas the second asks how much of the locally extractable work disappears after energy-basis dephasing. The step-by-step derivation of Eqs.~\eqref{eq:localpops}--\eqref{eq:WcohBi} is collected in Appendix~\ref{app:localcohinc}.

For the symmetric resonant even cat, set $m_i(t)=a|g(t)|^2$ and $r_i(t)=\exp[-2a(1-|g(t)|^2)]$. Equation~\eqref{eq:localpops} becomes
\begin{equation}
  p_n^{(i)}(t)=\frac{e^{-m_i}m_i^n}{n!\left(1+e^{-2a}\right)}\left[1+(-1)^n r_i\right].
  \label{eq:localcatpops}
\end{equation}
For the protocol used in the figures, $a=2.25$, so $0\le m_i\le a/3=0.75$. Appendix~\ref{app:localcohinc} gives an exact consecutive-ratio criterion for passivity throughout the charging branch. In the present family it reduces to
\begin{equation}
 \frac{a}{3}\tanh\!\left(\frac{2a}{3}\right)\le1,
 \qquad
 \frac{a}{6}\coth\!\left(\frac{2a}{3}\right)\le1.
 \label{eq:passivity-domain}
\end{equation}
The first inequality is the restrictive one and defines $a\le a_c\simeq3.098007$. Hence $a=2.25$ lies strictly inside the passive domain and
\begin{equation}
 p_0^{(i)}\ge p_1^{(i)}\ge p_2^{(i)}\ge\cdots
\end{equation}
at every time on the charging branch. The local diagonal state is therefore passive and
\begin{equation}
  \Wb_{B_i}^{\rm inc}(t)=0,\qquad \Wb_{B_i}^{\rm coh}(t)=\Wb_{B_i}(t)
  \label{eq:localallcoh}
\end{equation}
for each cell in the protocol studied in the figures. This statement concerns energetic coherence, not purity: $\nu_-^{(i)}$ can remain nonzero. Thus every unit of work extractable from one cell requires its energy-basis coherence, whereas the full battery develops $\Wb_B^{\rm inc}>0$ near complete transfer. Population-based work is therefore a genuinely collective feature in this protocol, invisible in each single-cell marginal.

\begin{figure*}[t]
  \centering
  \includegraphics[width=1\textwidth]{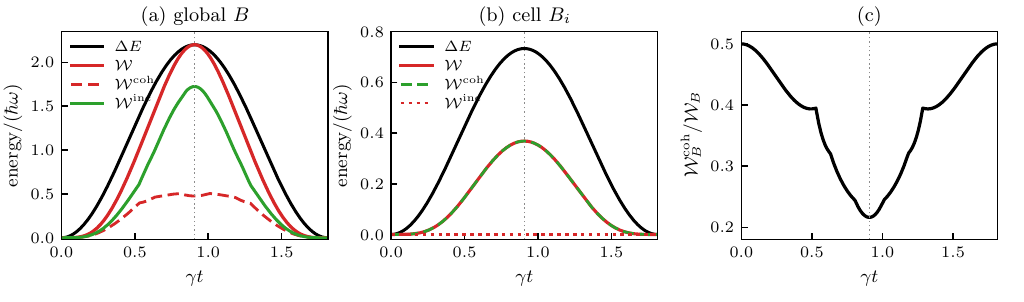}
  \caption{Global and local coherent--incoherent decomposition for the symmetric resonant even-cat protocol with $\alpha_0=1.5$, using the degeneracy-respecting spectral pinching of Eq.~\eqref{eq:dephasechannel}; the horizontal axis is $\gamma t$. (a) Global stored energy $\Delta E_B$ (black), total ergotropy $\Wb_B$ (solid red), coherent ergotropy $\Wb_B^{\rm coh}$ (dashed red), and incoherent ergotropy $\Wb_B^{\rm inc}$ (green). (b) Single-cell stored energy $\Delta E_{B_i}$ (black), with $\Wb_{B_i}$ and $\Wb_{B_i}^{\rm coh}$ superposed (solid red and dashed green); the dotted red line is $\Wb_{B_i}^{\rm inc}=0$. For compactness, the subsystem subscripts are suppressed in the legends inside panels (a) and (b). (c) Global coherent fraction $\Wb_B^{\rm coh}/\Wb_B$ (black). The local coherent fraction is not plotted separately because Eq.~\eqref{eq:localallcoh} gives $\Wb_{B_i}^{\rm coh}/\Wb_{B_i}=1$ whenever $\Wb_{B_i}>0$. At complete transfer, $\Wb_B^{\rm coh}/\Wb_B\simeq0.21625$ and $\Wb_B^{\rm inc}/\Wb_B\simeq0.78375$. The cusps in panel (c) arise from crossings in the ordered block eigenvalues $\{\pi_n\}$ entering the passive rearrangement and are therefore physical changes of ordering rather than numerical discontinuities. At the vacuum endpoints, where $\Wb_B=0$ and the ratio is undefined, panel (c) displays its continuous limiting value $1/2$. The vertical dotted line marks $t_*$.}
  \label{fig:cohinc}
\end{figure*}

Figure~\ref{fig:cohinc} compares the two decompositions. With the degeneracy-respecting pinching of Eq.~\eqref{eq:dephasechannel}, the global coherent fraction decreases during charging and reaches $\Wb_B^{\rm coh}/\Wb_B\simeq0.21625$ at $t_*$, while the local coherent fraction is identically one wherever $\Wb_{B_i}>0$. Thus approximately $78.4\%$ of the globally extractable work at complete transfer is population-based, even though every single-cell diagonal remains passive. The contrast strengthens the global--local distinction: population-based work emerges collectively while each marginal cell remains entirely coherence-supported.

\section{Concurrence, work, and the passive spectral penalty}\label{sec:penalty}

The extractable work of the full battery is lowered by the single spectral quantity $\nu_-$. We now connect that passive weight to entanglement and then resolve the multipartite correlations through exact Wootters concurrences on the two-branch supports.

\subsection{Concurrence, work, and power}\label{sec:poweridentity}

For the pure unitary bipartition $C|B$, the nonzero spectrum of $\rho_B$ is $\{\nu_+,\nu_-\}$. Both sides of the bipartition occupy supports of dimension at most two, so Wootters concurrence applies exactly after local orthonormalization of those supports (Appendix~\ref{app:tangles}). For this pure effective two-qubit state,
\begin{equation}
 \Cc_{C:B}=2\sqrt{\det\rho_B}=2\sqrt{\nu_+\nu_-}.
 \label{eq:CBpure}
\end{equation}
Hence
\begin{equation}
 \nu_-=\frac12\left(1-\sqrt{1-\Cc_{C:B}^2}\right),
\end{equation}
and substitution into Eq.~\eqref{eq:WB} gives
\begin{equation}
  \Wb_B(t)=\Delta E_B(t)-\frac{\hbar\omega_{\min}}{2}\Bigl(1-\sqrt{1-\Cc^2_{C:B}(t)}\Bigr).
  \label{eq:building}
\end{equation}
At fixed stored energy, $\Cc_{C:B}$ therefore lowers the globally extractable work monotonically. Unit global efficiency is equivalent to disentanglement from the charger whenever $\Delta E_B>0$,
\begin{equation}
  \eta_B=1\;\Longleftrightarrow\;\Cc_{C:B}=0,
  \qquad \Delta E_B>0.
  \label{eq:etaglobal}
\end{equation}

The same reasoning applies to a single cell through the pure bipartition $B_i|CB_jB_k$. Since $\rho_{B_i}$ has eigenvalues $\nu_\pm^{(i)}$,
\begin{equation}
 \Cc_{B_i:CB_jB_k}=2\sqrt{\nu_+^{(i)}\nu_-^{(i)}},
\end{equation}
and
\begin{equation}
  \Wb_{B_i}=\Delta E_{B_i}-\frac{\hbar\omega_i}{2}\Bigl(1-\sqrt{1-\Cc^2_{B_i:CB_jB_k}}\Bigr).
  \label{eq:WBilocal}
\end{equation}
Thus, for $\Delta E_{B_i}>0$, unit local efficiency is equivalent to $\Cc_{B_i:CB_jB_k}=0$, i.e. to purity of $B_i$ relative to its complement. For nonzero superposition coefficients this is equivalent to
\begin{equation}
 (1-|q_i|^2)(1-|sQ_i|^2)=0.
\end{equation}
Purity can arise in either of two ways: the two local branches coincide, $|q_i|=1$, or the complementary branches coincide, $|sQ_i|=1$.

The spectral connection also extends to the rate of extractable work. With $P_E=\dot{\Delta E}_B$ and $P_{\Wb}=\dot{\Wb}_B$, differentiating Eq.~\eqref{eq:building} gives
\begin{equation}
  P_{\Wb}=P_E-\frac{\hbar\omega_{\min}}{2}\,
  \frac{\Cc_{C:B}\dot{\Cc}_{C:B}}{\sqrt{1-\Cc^2_{C:B}}}.
  \label{eq:power}
\end{equation}
Equivalently, $P_{\Wb}=P_E-\hbar\omega_{\min}\dot\nu_-$. When $\dot\nu_->0$, incoming energy partly increases passive spectral weight; when $\dot\nu_-<0$, spectral purification enhances the ergotropic power relative to the energy power.

\subsection{Residual entanglement and monogamy redistribution}\label{sec:residual}

All bipartite concurrences used below are Wootters concurrences~\cite{Wootters1998} evaluated on exact effective two-qubit supports. The squared-concurrence residuals follow the CKW monogamy construction~\cite{CKW2000}, together with its general multiqubit monogamy extension~\cite{Osborne2006}. For a subset $X$ of parties, let $q_X$ denote the overlap between its two branch states; thus $q_C=s$, $q_{B_i}=q_i$, and overlaps multiply for composite parties. For a bipartition $A|B$ obtained by tracing a remainder $R$, Appendix~\ref{app:tangles} gives the unitary master formula
\begin{equation}
  \Cc_{A:B}=2|\mathcal N_0|^2|c_\alpha c_\beta|\,|q_R|
  \sqrt{1-|q_A|^2}\sqrt{1-|q_B|^2}.
  \label{eq:Cgeneral}
\end{equation}
This expression applies whether the reduced state $\rho_{AB}$ is pure or mixed. Setting
\begin{equation}
 x_C=|s|^2,\qquad x_i=|q_i|^2,\qquad
 \mathcal A=4|\mathcal N_0|^4|c_\alpha|^2|c_\beta|^2,
 \label{eq:xdefs}
\end{equation}
we obtain
\begin{align}
  \Cc^2_{B_i:CB_jB_k}&=\mathcal A(1-x_i)(1-x_Cx_jx_k),\label{eq:Conetangle}\\
  \Cc^2_{B_i:B_jB_k}&=\mathcal A x_C(1-x_i)(1-x_jx_k),\label{eq:Cintermediate}\\
  \Cc^2_{B_i:B_j}&=\mathcal A x_Cx_k(1-x_i)(1-x_j),\label{eq:Cpair}\\
  \Cc^2_{B_i:C}&=\mathcal A x_jx_k(1-x_i)(1-x_C).\label{eq:CpairC}
\end{align}

The battery-internal CKW monogamy residual associated with focus cell $B_i$ is
\begin{align}
 \mathcal R_i^{(B)}
 &\equiv \Cc^2_{B_i:B_jB_k}-\Cc^2_{B_i:B_j}-\Cc^2_{B_i:B_k}\nonumber\\
 &=\mathcal A x_C(1-x_i)(1-x_j)(1-x_k).
 \label{eq:RBresidual}
\end{align}
When the charger is entangled with the battery, $\rho_{B_1B_2B_3}$ is generally mixed. Accordingly, $\mathcal R_i^{(B)}$ is a CKW monogamy residual and is not identified here with the convex-roof mixed-state three-tangle. Its closed expression is symmetric under every permutation of the cells, so within the present two-branch family it is independent of the chosen focus cell. Whenever the battery is pure, the usual pure-state CKW relation applies and
\begin{equation}
 \mathcal R_i^{(B)}=\tau_3(B_1|B_2|B_3).
 \label{eq:RBequalstau}
\end{equation}

A second natural tripartition is $B_i|C|B_jB_k$. During unitary evolution the global state is pure and each of these three parties has support dimension at most two. Hence its CKW residual is a genuine pure-state three-tangle,
\begin{align}
 \mathcal R_i^{(C)}
 &\equiv \Cc^2_{B_i:CB_jB_k}-\Cc^2_{B_i:B_jB_k}-\Cc^2_{B_i:C}\nonumber\\
 &=\mathcal A(1-x_i)(1-x_C)(1-x_jx_k).
 \label{eq:RCresidual}
\end{align}
Thus $\mathcal R_i^{(C)}=\tau_3(B_i|C|B_jB_k)$ in the unitary problem.

The two residuals provide a convenient exact resolution of the one-focus four-party monogamy score,
\begin{align}
 \mathcal R_i^{(4)}&\equiv \mathcal R_i^{(B)}+\mathcal R_i^{(C)}\nonumber\\
 &=\Cc^2_{B_i:CB_jB_k}-\Cc^2_{B_i:B_j}-\Cc^2_{B_i:B_k}-\Cc^2_{B_i:C}.
 \label{eq:monogamy}
\end{align}
The factors $x_C$ and $1-x_C$ make the redistribution transparent. Branch distinguishability retained by the charger supports $\mathcal R_i^{(C)}$; when the charger branches coalesce, $x_C\to1$, $\mathcal R_i^{(C)}\to0$, while the battery-internal residual can remain finite. The same branch overlaps that set the passive spectral penalty therefore also determine where the CKW residual correlations reside.

\section{Optimal charging in the symmetric resonant even-cat protocol}\label{sec:symmetric}

We now specialize to the symmetric resonant regime of Eq.~\eqref{eq:fg} and to an even-cat charger~\cite{GerryKnight1997},
\begin{equation}
  c_\alpha=c_\beta=\tfrac{1}{\sqrt2},\qquad \beta_0=-\alpha_0,\qquad\alpha_0\in\mathbb R,
  \label{eq:evencat}
\end{equation}
with $a\equiv|\alpha_0|^2$ and mean initial excitation number $\bar n_0=a\tanh a$. The normalization is $|\mathcal N_0|^2=(1+e^{-2a})^{-1}$, so $\mathcal A=(1+e^{-2a})^{-2}$. The even cat maximizes the branch separation at fixed $|\alpha_0|$ and makes all overlap moduli elementary functions of the transferred fraction.

\subsection{Exact optimality at full transfer}\label{sec:optimalpoint}

The stored energy is
\begin{equation}
 \Delta E_B(t)=\hbar\omega\bar n_0\Theta(t)
 =\hbar\omega\bar n_0\sin^2(\sqrt3\gamma t).
\end{equation}
At $t_*=\pi/(2\sqrt3\gamma)$, $\Theta(t_*)=1$ and $\Delta E_B(t_*)=\hbar\omega\bar n_0$, the maximum of the first charging branch. Since $f(t_*)=0$, the two charger branches coincide at the vacuum and $s(t_*)=1$. Equations~\eqref{eq:detB} and~\eqref{eq:CBpure} then give
\begin{equation}
 \Cc_{C:B}(t_*)=0,\qquad \nu_-(t_*)=0,
\end{equation}
and therefore
\begin{equation}
  \Wb_B(t_*)=\Delta E_B(t_*),\qquad\eta_B(t_*)=1.
  \label{eq:optimal}
\end{equation}
Appendix~\ref{app:mono} proves independently that $\Wb_B$ is nondecreasing in $\Theta$ and strictly increasing for $0<\Theta\le1$, so this point is also the first global ergotropy maximum.

\begin{figure}[t]
  \centering
  \includegraphics[width=\columnwidth]{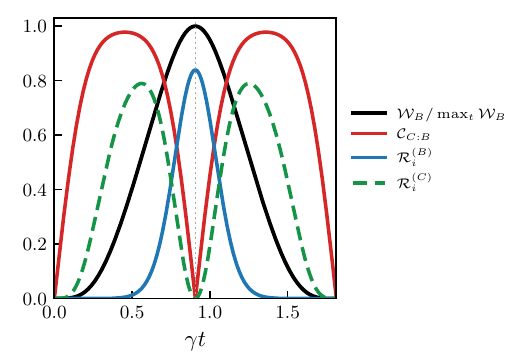}
  \caption{Symmetric resonant protocol for $\alpha_0=1.5$, $\beta_0=-1.5$, and $c_\alpha=c_\beta=1/\sqrt2$, plotted against $\gamma t$. The black curve is $\Wb_B/\max_t\Wb_B$, the solid red curve is $\Cc_{C:B}$, the blue curve is the battery-internal CKW residual $\mathcal R_i^{(B)}$, and the dashed green curve is the charger-involving residual $\mathcal R_i^{(C)}$. At $t_*$, $\Cc_{C:B}$ and $\mathcal R_i^{(C)}$ vanish, while $\mathcal R_i^{(B)}$ is maximal and, because the battery is pure there, coincides with its standard three-tangle.}
  \label{fig:symmetric}
\end{figure}

The optimal point is not a point at which all correlations disappear. In the even-cat protocol,
\begin{equation}
 x_C(t)=e^{-4a\Xi(t)},\qquad x(t)\equiv x_i(t)=e^{-4a\Theta(t)/3}.
\end{equation}
Thus
\begin{align}
 \mathcal R_i^{(B)}&=\mathcal A x_C(1-x)^3,\\
 \mathcal R_i^{(C)}&=\mathcal A(1-x)(1-x_C)(1-x^2).
\end{align}
At full transfer,
\begin{equation}
  \mathcal R_i^{(C)}(t_*)=0,\qquad
  \mathcal R_i^{(B)}(t_*)=\mathcal A\bigl(1-e^{-4a/3}\bigr)^3.
  \label{eq:Rstar}
\end{equation}
For every nontrivial even cat $a>0$, the second quantity is strictly positive; it vanishes only at the trivial endpoint $a=0$. Since $\rho_B(t_*)$ is pure, Eq.~\eqref{eq:RBequalstau} identifies $\mathcal R_i^{(B)}(t_*)$ with the standard three-tangle of the three battery cells.

The same $t_*$ is not locally optimal in the sense of unit efficiency. Although $\Delta E_{B_i}=\Delta E_B/3$, a single cell remains entangled with the other cells at full transfer, so $\nu_-^{(i)}(t_*)$ is generally nonzero. For $\alpha_0=1.5$, one finds $\nu_-^{(i)}(t_*)\simeq0.36504$ and $\eta_{B_i}(t_*)\simeq0.50234$.

The timing of the local ergotropy maximum can nevertheless be established exactly. Let $y\equiv\Theta\in[0,1]$. Appendix~\ref{app:mono} gives
\begin{equation}
 \frac{1}{\hbar\omega}\frac{d\Wb_{B_i}}{dy}
 =\frac{a}{3\cosh a}
 \left[
 \sinh a-\sinh\!\left(a\left(1-\frac{2y}{3}\right)\right)
 \right].
 \label{eq:localmono}
\end{equation}
For $a>0$ this derivative vanishes at $y=0$ and is strictly positive for every $0<y\le1$. Therefore
\begin{equation}
 \arg\max_{0\le y\le1}\Wb_{B_i}(y)=1,
\end{equation}
so the first local and global ergotropy maxima occur at the same $t_*$. What differs is their efficiency: the collective battery reaches unity, whereas each cell remains mixed and locally suboptimal. For $a=2.25$, Eq.~\eqref{eq:localallcoh} adds the stronger statement that the entire single-cell ergotropy is coherence-supported.

\begin{figure}[t]
  \centering
  \includegraphics[width=\columnwidth]{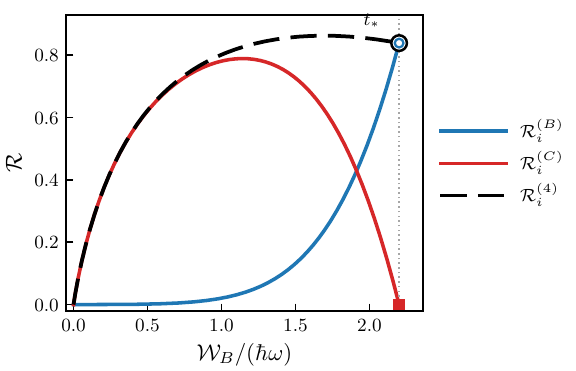}
  \caption{CKW monogamy residuals as functions of the global ergotropy in the symmetric resonant even-cat protocol. The blue curve is $\mathcal R_i^{(B)}$, the red curve is $\mathcal R_i^{(C)}$, and the black dashed curve is $\mathcal R_i^{(4)}=\mathcal R_i^{(B)}+\mathcal R_i^{(C)}$. At $t_*$ the global ergotropy is maximal, $\mathcal R_i^{(C)}$ vanishes, and the remaining residual is internal to the battery and equals its pure-state three-tangle.}
  \label{fig:residual}
\end{figure}

\begin{figure*}[t]
  \centering
  \includegraphics[width=0.75\textwidth]{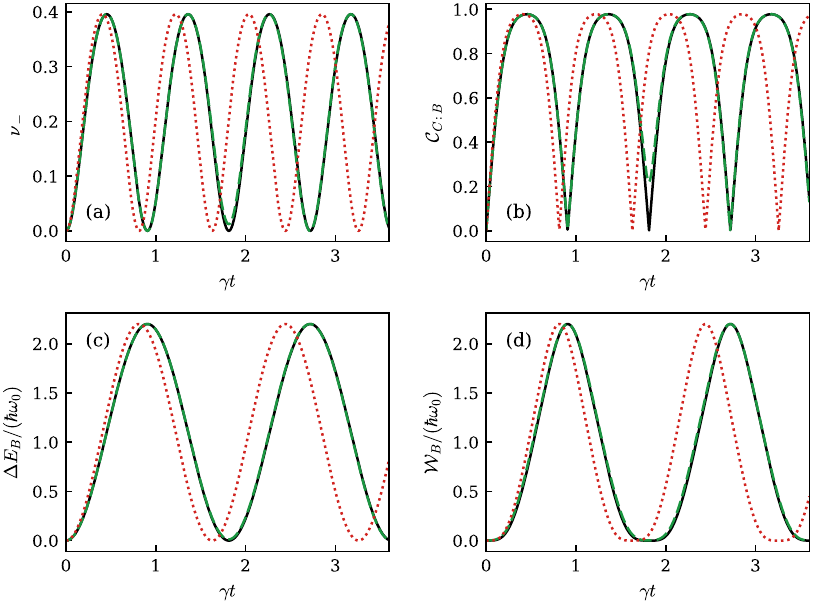}
  \caption{Persistence of the rank-two relations away from the symmetric point, plotted against the reference scaled time $\gamma t$ for $\alpha_0=1.5$. Panels show (a) $\nu_-$, (b) $\Cc_{C:B}$, (c) $\Delta E_B/(\hbar\omega_0)$, and (d) $\Wb_B/(\hbar\omega_0)$. Black solid curves are the symmetric resonant reference; green dashed curves use detuned cells $\omega_i=(0.92,1.00,1.08)\omega_0$ at uniform coupling; red dotted curves use resonant cells with $\gamma_i=(0.4,1.0,1.6)\gamma$. The identity $\Wb_B=\Delta E_B-\hbar\omega_{\min}\nu_-$ holds pointwise in all three cases; the coincidence of extrema is not asserted away from the symmetric protocol.}
  \label{fig:robust}
\end{figure*}

The rank-two spectral identities do not rely on resonance or equal couplings, but the exact timing statements above do. For arbitrary $\omega_i$ and $\gamma_i$, the amplitudes follow from Eq.~\eqref{eq:M} and the identities $\Wb_B=\Delta E_B-\hbar\omega_{\min}\nu_-$ and Eq.~\eqref{eq:Cgeneral} remain exact. The coincidence of the corresponding extrema, however, must be determined from the actual dynamics. Figure~\ref{fig:robust} illustrates two perturbations: detuned cells $\omega_i\in\{0.92,1.00,1.08\}\omega_0$ at uniform coupling and resonant cells with $\gamma_i\in\{0.4,1.0,1.6\}\gamma$. In both examples the rank-two relation survives pointwise, while the times and depths of high-extractability events are deformed.

\subsection{Power-resolved redistribution}\label{sec:powerresolved}

In the symmetric resonant regime,
\begin{align}
  P_E(t)&=\hbar\omega\bar n_0\sqrt3\,\gamma\sin(2\sqrt3\gamma t),\nonumber\\
  P_{\Wb}(t)&=P_E(t)-\hbar\omega\dot\nu_-(t),
  \label{eq:powersym}
\end{align}
and Eq.~\eqref{eq:power} rewrites the second term through $\Cc_{C:B}\dot{\Cc}_{C:B}$. With $y=\Theta$, the residuals are
\begin{align}
 \mathcal R_i^{(B)}(y)&=\mathcal A x_C(y)[1-x(y)]^3,\\
 \mathcal R_i^{(C)}(y)&=\mathcal A[1-x(y)][1-x_C(y)][1-x^2(y)],
\end{align}
where $x_C(y)=e^{-4a(1-y)}$ and $x(y)=e^{-4ay/3}$. Early in the cycle, charger branch distinguishability supports $\mathcal R_i^{(C)}$; near full transfer, the charger branches coalesce and the CKW residual becomes entirely battery-internal.

\begin{figure*}[t]
  \centering
  \includegraphics[width=0.95\textwidth]{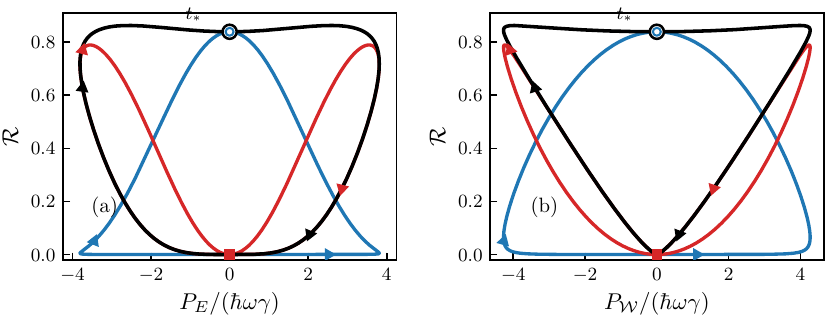}
  \caption{CKW monogamy residuals against instantaneous power, normalized by $\hbar\omega\gamma$. Blue is $\mathcal R_i^{(B)}$, red is $\mathcal R_i^{(C)}$, and black is $\mathcal R_i^{(4)}$. (a) Energy power $P_E$. (b) Ergotropic power $P_{\Wb}$, which additionally contains the spectral-purification term $-\hbar\omega\dot\nu_-$. Arrowheads indicate the direction of time.}
  \label{fig:power}
\end{figure*}

Figure~\ref{fig:power} displays these residuals against instantaneous energy and ergotropic power. When $\dot\Cc_{C:B}>0$, charger-battery entanglement grows and $P_{\Wb}<P_E$; when $\dot\Cc_{C:B}<0$, spectral purification gives $P_{\Wb}>P_E$. The apparent singularity of Eq.~\eqref{eq:power} at $\Cc_{C:B}=1$ is removable. Indeed, differentiating $\Cc_{C:B}^2=4\nu_+\nu_-$ and using $\nu_+=1-\nu_-$ gives
\begin{equation}
 \frac{\Cc_{C:B}\dot{\Cc}_{C:B}}{\sqrt{1-\Cc_{C:B}^2}}
 =2\dot\nu_-,
 \label{eq:power-regularity}
\end{equation}
At $\Cc_{C:B}=1$ the equality is understood by continuity, so the spectral-purification term remains finite whenever $\dot\nu_-$ is finite. The resulting deformation between energy-power and ergotropic-power representations resolves the same finite-size redistribution encoded by the branch overlaps and CKW residuals.

\section{Single-photon loss}
\label{sec:diss}

We now include zero-temperature single-photon loss,
\begin{equation}
  \dot\rho = -\frac{i}{\hbar}[\hat H,\rho]
  + \sum_{\mu\in\{c,1,2,3\}} \kappa_\mu
    \Bigl( \hat a_\mu \rho\, \hat a_\mu^\dagger
    - \tfrac12 \{ \hat a_\mu^\dagger \hat a_\mu , \rho \} \Bigr),
  \label{eq:lindblad}
\end{equation}
with $\hat a_c=\hat c$ and $\hat a_i=\hat b_i$. Linear damping preserves the two-branch coherent-dyad structure. Writing $\ket{\bm z}$ for a normalized product coherent state, the solution with initial condition~\eqref{eq:init} has the form
\begin{equation}
  \rho(t) = |\mathcal N_0|^2 \!\!\sum_{A,B\in\{\alpha,\beta\}}\!\!
  c_A c_B^{*}\, w_{AB}(t)\, |\bm z^A(t)\rangle\langle \bm z^B(t)|,
  \label{eq:rhodiss}
\end{equation}
where
\begin{equation}
 \dot{\bm z}^A=-(iM+\Lambda)\bm z^A,
 \qquad
 \Lambda=\tfrac12\operatorname{diag}(\kappa_c,\kappa_1,\kappa_2,\kappa_3),
\end{equation}
$w_{\alpha\alpha}=w_{\beta\beta}=1$, and $w_{\alpha\beta}=w_{\beta\alpha}^*=e^{-\Gamma(t)}$, with
\begin{equation}
  \Gamma(t) = \frac12 \int_0^t ds \sum_\mu \kappa_\mu
  \Bigl[ |z^\alpha_\mu|^2 + |z^\beta_\mu|^2
  - 2 z^\alpha_\mu z^{\beta *}_\mu \Bigr].
  \label{eq:Gamma}
\end{equation}
The derivation is given in Appendix~\ref{app:diss}. We define the environment-overlap factor
\begin{equation}
 x_E\equiv |w_{\alpha\beta}|^2=e^{-2\operatorname{Re}\Gamma},
 \label{eq:xE}
\end{equation}
and retain $x_C=|s|^2$, $x_i=|q_i|^2$, and $x_B=|q_B|^2$.

Every reduced state remains rank at most two. The battery determinants become
\begin{align}
  \det \rho_B &= |\mathcal N_0|^4 |c_\alpha|^2 |c_\beta|^2
  (1 - x_E x_C)(1 - x_B),\label{eq:detdissB}\\
  \det \rho_{B_i} &= |\mathcal N_0|^4 |c_\alpha|^2 |c_\beta|^2
  (1 - x_E x_Cx_jx_k)(1 - x_i).
  \label{eq:detdiss}
\end{align}
Thus the spectral ergotropy formulas remain exact,
\begin{align}
 \Wb_B&=\Delta E_B-\hbar\omega_{\min}\nu_-,\\
 \Wb_{B_i}&=\Delta E_{B_i}-\hbar\omega_i\nu_-^{(i)},
\end{align}
but the smaller eigenvalues are no longer determined by entanglement alone because the global state of charger plus battery is mixed.

For any normalized rank-two state, define
\begin{equation}
 \mathcal M_B\equiv2\sqrt{\nu_+\nu_-}
 =\sqrt{2(1-\operatorname{Tr}\rho_B^2)},
\end{equation}
and analogously $\mathcal M_{B_i}$. The exact Wootters formula of Appendix~\ref{app:tangles} gives
\begin{equation}
 \Cc_{C:B}^2=\mathcal A x_E(1-x_C)(1-x_B),
 \label{eq:CdissCB}
\end{equation}
whereas Eq.~\eqref{eq:detdissB} yields
\begin{equation}
 \mathcal M_B^2=\mathcal A(1-x_Ex_C)(1-x_B).
\end{equation}
The difference is therefore the exact identity
\begin{equation}
 \boxed{\mathcal M_B^2=\Cc_{C:B}^2+\mathcal A(1-x_E)(1-x_B).}
 \label{eq:mixednesssplit}
\end{equation}
The first term is the squared charger-battery concurrence; the second is the additional mixedness created by which-branch information lost to the environment. Similarly,
\begin{align}
 \Cc_{B_i:CB_jB_k}^2&=\mathcal A x_E(1-x_i)(1-x_Cx_jx_k),\\
 \mathcal M_{B_i}^2&=\Cc_{B_i:CB_jB_k}^2+\mathcal A(1-x_E)(1-x_i).
 \label{eq:mixednesssplitlocal}
\end{align}
Equations~\eqref{eq:mixednesssplit} and~\eqref{eq:mixednesssplitlocal} reduce to the unitary concurrence identities at $x_E=1$.

The ergotropies can consequently be written as
\begin{align}
  \Wb_B &= \Delta E_B - \frac{\hbar\omega_{\min}}{2}
  \left(1-\sqrt{1-\mathcal M_B^2}\right),\\
  \Wb_{B_i} &= \Delta E_{B_i} - \frac{\hbar\omega_i}{2}
  \left(1-\sqrt{1-\mathcal M_{B_i}^2}\right),
\end{align}
and
\begin{equation}
  P_{\Wb}=P_E-\frac{\hbar\omega_{\min}}{2}
  \frac{\mathcal M_B\dot{\mathcal M}_B}{\sqrt{1-\mathcal M_B^2}}.
  \label{eq:Power_diss}
\end{equation}
Only in the unitary limit $x_E=1$ can the second term be interpreted purely as an entanglement-rate contribution.

For $\Delta E_B>0$, unit efficiency requires $\mathcal M_B=0$. With nonzero superposition coefficients this occurs when
\begin{equation}
 x_B=1\qquad\text{or}\qquad x_E x_C=1.
 \label{eq:disspureconditions}
\end{equation}
Under genuine loss, $x_E<1$ once branch information has leaked to the environment, so the second condition cannot be met. In the nontrivial even-cat charging branch, $x_B<1$ as soon as energy has entered the battery. Therefore $\eta_B<1$ at all nontrivial charging times with $\Delta E_B>0$ in that protocol.

\subsection{Uniform loss}\label{sec:diss:uniform}

For $\kappa_\mu=\kappa$, damping commutes with the star dynamics,
\begin{equation}
 \bm z^A(t)=e^{-\kappa t/2}e^{-iMt}\bm z^A(0).
\end{equation}
For the even cat, $\Gamma(t)=2a(1-e^{-\kappa t})$ is real. Define
\begin{equation}
  \widetilde\Theta(t)=e^{-\kappa t}\Theta(t)
  =e^{-\kappa t}\sin^2(\sqrt3\gamma t).
  \label{eq:Thetatilde}
\end{equation}
Then
\begin{align}
 x_Ex_C&=e^{-4a(1-\widetilde\Theta)},\\
 x_i&=e^{-4a\widetilde\Theta/3},\\
 x_B&=e^{-4a\widetilde\Theta}.
 \label{eq:loss-battery-overlaps}
\end{align}
Consequently, battery-only spectral quantities and Fock-diagonal populations have exactly the lossless functional form after $y=\Theta\to\widetilde\Theta$. This includes $\det\rho_B$, $\det\rho_{B_i}$, $\Delta E_B$, $\Wb_B$, $\Wb_{B_i}$, and the population formulas of Sec.~\ref{sec:cohinc}. In particular,
\begin{equation}
 \Delta E_B=\hbar\omega\bar n_0\widetilde\Theta.
\end{equation}
For $a=2.25$, the local passivity proof remains valid because $0\le\widetilde\Theta\le1$, so $\Wb_{B_i}^{\rm inc}=0$ also under uniform loss. Equation~\eqref{eq:localmono} likewise implies that $\Wb_{B_i}$ is increasing as a function of $\widetilde\Theta$.

Charger-involving correlations require $x_E$ and $x_C$ separately:
\begin{equation}
 x_E=e^{-4a(1-e^{-\kappa t})},\qquad
 x_C=e^{-4ae^{-\kappa t}[1-\Theta(t)]}.
 \label{eq:xExCuniform}
\end{equation}
Thus neither $\Cc_{C:B}$ nor $\mathcal R_i^{(C)}$ is obtained from its lossless curve by the replacement $y\to\widetilde\Theta$ alone.

Since $\Delta E_B$, $\Wb_B$, and $\Wb_{B_i}$ are increasing functions of $\widetilde\Theta$ on the charging branch, their maxima occur at the maximum of $\widetilde\Theta$. Differentiating Eq.~\eqref{eq:Thetatilde} gives
\begin{align}
  \sqrt3\gamma\, t_{\mathrm{opt}} &= \arctan \upsilon,\\
  \widetilde\Theta_{\max} &= e^{-\kappa t_{\mathrm{opt}}}\frac{\upsilon^2}{1+\upsilon^2},\\
  \upsilon &= \frac{2\sqrt3\gamma}{\kappa}.
  \label{eq:topt}
\end{align}
Hence $t_{\mathrm{opt}}<t_*$ and $t_{\mathrm{opt}}\to t_*$ as $\kappa\to0$.

At the nominal full-transfer time, the charger amplitudes still vanish and therefore $x_C(t_*)=1$. Consequently,
\begin{equation}
 \Cc_{C:B}(t_*)=0,
 \end{equation}
but $x_E(t_*)<1$ for $\kappa>0$, so Eq.~\eqref{eq:mixednesssplit} gives $\mathcal M_B(t_*)>0$ whenever $x_B(t_*)<1$. Charger disentanglement therefore no longer implies battery purification. For weak loss,
\begin{equation}
 1-\eta_B(t_*)=\kappa t_*+\mathcal O((\kappa t_*)^2),
 \label{eq:etaweakloss}
\end{equation}
and the leading coefficient is independent of the cat amplitude.

\subsection{Entanglement redistribution under loss}\label{sec:diss:entanglement}

The effective-support argument remains exact under loss. For any bipartition $A|B$ obtained by tracing a remainder $R$, Appendix~\ref{app:tangles} gives
\begin{equation}
 \Cc_{A:B}^2
 =\mathcal A x_E x_R(1-x_A)(1-x_B),
 \label{eq:CgeneralLoss}
\end{equation}
where $x_R=|q_R|^2$. Thus no Fock-space truncation is required for the dissipative Wootters concurrences either.

The corresponding CKW residuals are
\begin{align}
 \mathcal R_i^{(B)}
 &=\mathcal A x_E x_C(1-x_i)(1-x_j)(1-x_k),\label{eq:RBloss}\\
 \mathcal R_i^{(C)}
 &=\mathcal A x_E(1-x_i)(1-x_C)(1-x_jx_k).
 \label{eq:RCloss}
\end{align}
Both residuals are manifestly nonnegative. For $x_E<1$, the retained three-party states are generically mixed whenever their two branch states remain linearly independent; hence Eqs.~\eqref{eq:RBloss} and~\eqref{eq:RCloss} are, in the generic dissipative regime, mixed-state CKW monogamy residuals rather than pure-state three-tangles. Their sum is the one-focus score $\mathcal R_i^{(4)}$ of Eq.~\eqref{eq:monogamy}, with each squared concurrence replaced by its dissipative value.

For uniform loss, $\mathcal R_i^{(B)}$ depends only on $\widetilde\Theta$ because $x_Ex_C=e^{-4a(1-\widetilde\Theta)}$; it is increasing over the charging branch and therefore peaks at $t_{\mathrm{opt}}$. By contrast, $\mathcal R_i^{(C)}$ depends separately on $x_E$ and $x_C$ and vanishes at $t_*$ because $x_C(t_*)=1$.

\begin{figure}[t]
  \centering
  \includegraphics[width=\columnwidth]{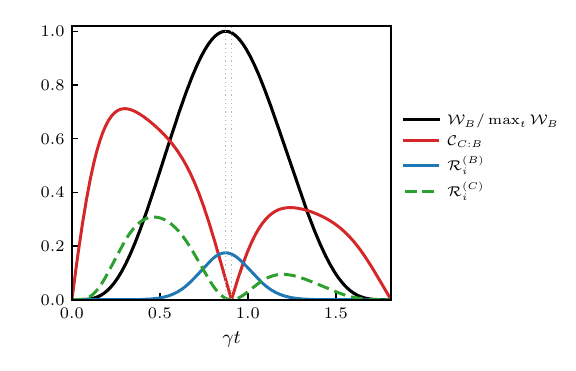}
  \caption{Lossy counterpart of Fig.~\ref{fig:symmetric} for uniform loss $\kappa=0.2\gamma$. The black curve is $\Wb_B/\max_t\Wb_B$, the solid red curve is the mixed-state Wootters concurrence $\Cc_{C:B}$, the blue curve is the battery-internal residual $\mathcal R_i^{(B)}$, and the dashed green curve is the charger-involving residual $\mathcal R_i^{(C)}$. The ergotropy and $\mathcal R_i^{(B)}$ peak at $t_{\rm opt}$, whereas $\Cc_{C:B}$ and $\mathcal R_i^{(C)}$ vanish at $t_*$.}
  \label{fig:losstangles}
\end{figure}

\begin{figure}[t]
  \centering
  \includegraphics[width=\columnwidth]{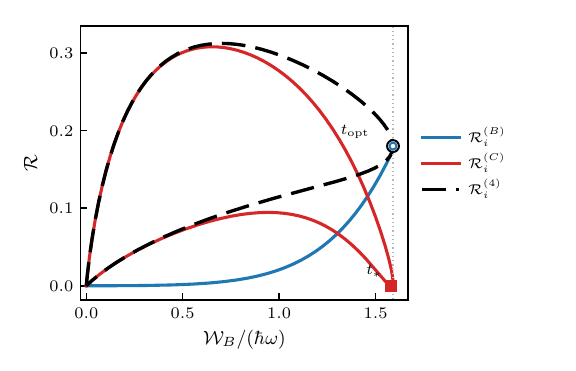}
  \caption{Dissipative CKW monogamy residuals as functions of the global ergotropy for $\kappa=0.2\gamma$. The blue curve is $\mathcal R_i^{(B)}$, the red curve is $\mathcal R_i^{(C)}$, and the black dashed curve is $\mathcal R_i^{(4)}$. The circle marks $t_{\rm opt}$ and the square marks $t_*$. Loss separates maximal battery-only extraction from charger factorization because the environment retains branch information.}
  \label{fig:lossresidual}
\end{figure}

The two characteristic times now have distinct meanings. At $t_{\mathrm{opt}}$, battery-only quantities controlled by $\widetilde\Theta$ are maximal. At $t_*$, the charger factorizes from the battery, so $\Cc_{C:B}=\mathcal R_i^{(C)}=0$, but Eq.~\eqref{eq:mixednesssplit} shows that the battery remains mixed because $x_E<1$. The missing purity is precisely the contribution associated with which-branch information stored in the environment.

\section{Discussion and conclusions}\label{sec:disc}

We have derived an exact rank-two description of work extraction and correlation redistribution in a non-Gaussian bosonic quantum battery. The physical Hilbert spaces remain infinite dimensional and can populate arbitrarily many Fock levels; the reduction occurs only at the level of the dynamically generated two-branch support. This distinction is essential: the oscillator dynamics is solved without truncation, while exact local isometries subsequently map each relevant branch span to an effective qubit for spectral and entanglement calculations.

Under unitary dynamics, the passive spectral penalty of the reduced battery is determined exactly by the charger-battery concurrence. Equation~\eqref{eq:building} relates the global ergotropy to the Wootters concurrence $\Cc_{C:B}$, and Eq.~\eqref{eq:power} relates ergotropic power to its rate of change. The corresponding single-cell relation involves $\Cc_{B_i:CB_jB_k}$. These identities rely on purity of the corresponding global bipartitions and therefore must not be carried unchanged into the dissipative problem.

The multipartite analysis is naturally phrased in terms of CKW monogamy residuals. The battery-internal quantity $\mathcal R_i^{(B)}$ is generally a residual of a mixed three-cell state and is not identified with the convex-roof mixed-state three-tangle. Whenever the battery is pure, in particular at complete transfer in the unitary symmetric protocol, it coincides with the standard three-tangle. The charger-involving $\mathcal R_i^{(C)}$ is a genuine three-tangle of the effective pure tripartition $B_i|C|B_jB_k$ during unitary evolution. This distinction removes an ambiguity between monogamy scores and genuine multipartite entanglement monotones while preserving the exact overlap formulas that expose the redistribution.

For the symmetric resonant even-cat protocol, complete transfer has a particularly rigid structure. At $t_*$ the charger disentangles, the full battery is pure, $\eta_B=1$, and $\mathcal R_i^{(B)}$ is finite for every nontrivial cat. The individual cells nevertheless remain mixed. We proved that their ergotropy is strictly increasing with the transferred fraction for every $a>0$ on the charging branch, so the first local and global ergotropy maxima occur at the same time even though their efficiencies differ. For the parameter used in the figures, $a=2.25$, the single-cell Fock diagonal is passive throughout the charging branch; hence $\Wb_{B_i}^{\rm inc}=0$ and all locally extractable work is coherence-supported, while the full battery acquires a finite incoherent contribution.

Under single-photon loss, the two-branch rank-two structure survives while the passive eigenvalue changes meaning. The exact identity
\begin{equation}
 \mathcal M_B^2=\Cc_{C:B}^2+\mathcal A(1-x_E)(1-x_B)
\end{equation}
separates the mixedness produced by charger-battery entanglement from that generated by information leaked to the environment. Hence charger factorization at $t_*$ no longer guarantees a pure battery. Under uniform loss, battery-only spectral quantities remain functions of $\widetilde\Theta$ and peak at $t_{\rm opt}<t_*$. Charger-involving correlations instead depend separately on $x_E$ and $x_C$ and vanish at $t_*$. Thus loss splits the coincidence between disentanglement and optimal extraction in a quantitatively controlled way.

Whenever the evolution preserves two coherent branches, the rank-two spectral construction and the Wootters formula extend to any number of battery modes. Multipartite residuals, however, cannot automatically be interpreted as genuine $N$-partite monotones; such an interpretation requires a definition adapted to the partition and mixedness structure. The coincidence of local and global optimal times is likewise proved only for the symmetric resonant even-cat protocol. Away from that point, the spectral identities remain exact, but the extrema must be obtained from the actual dynamics.

Bosonic platforms supporting coherent-state superpositions and beam-splitter-like couplings provide natural candidate settings for realizing the model, including circuit and cavity QED and multimode photonic architectures. For a few modes, tomography of the reduced supports could directly test the concurrence--ergotropy identities, CKW-residual redistribution, and environmental mixedness decomposition. Measurement-assisted work extraction~\cite{Francica2017daemonic,Morrone2023daemonic,Pushpan2025} is a natural next step: the gap between collective and independent extraction identifies inaccessible work, while a daemonic advantage requires a separate optimization over measurements and conditional operations.

\begin{acknowledgments}
A.C.S.C. acknowledges support from CNPq/Brazil under Grant No. 308730/2023-2; J.G.G.O.Jr. acknowledges financial support from the Brazilian agency CAPES under Grant No.\ 88887.909640/2023-00.
\end{acknowledgments}

\section*{Data Availability}

No external datasets were used in this theoretical work.
All figures and numerical values reported in the manuscript can be reproduced directly from the analytical expressions given in the main text and appendices.

\bibliography{references}

\appendix

\section{Coherent-state propagation}\label{app:prop}
This appendix summarizes the coherent-state propagation used in Eq.~\eqref{eq:psit}. Define the vector of annihilation operators
\[
  \hat{\bm a}=(\hat c,\hat b_1,\hat b_2,\hat b_3)^T .
\]
From the Hamiltonian~\eqref{eq:H}, the Heisenberg equations are linear,
\begin{equation}
  i\,\frac{d}{dt}\hat{\bm a}(t)=M\hat{\bm a}(t),
\end{equation}
with the single-particle matrix $M$ given in Eq.~\eqref{eq:M}. Hence
\begin{equation}
  \hat{\bm a}(t)=e^{-iMt}\hat{\bm a}(0).
  \label{eq:app-prop-operator}
\end{equation}
This equation is exact for arbitrary frequencies $\omega_i$ and couplings $\gamma_i$. The number-conserving form of the Hamiltonian ensures that no creation operator appears in Eq.~\eqref{eq:app-prop-operator}; therefore, the evolution maps coherent products into coherent products rather than into squeezed Gaussian states.

To see this explicitly, consider the product coherent state
\begin{equation}
 \ket{\bm z(0)}=\ket{z_c(0)}_C\bigotimes_i\ket{z_i(0)}_{B_i}.
 \label{eq:app-product-coherent}
\end{equation}
It is a joint eigenstate of all annihilation operators. Equation~\eqref{eq:app-prop-operator} therefore implies that the evolved state remains a joint eigenstate, with amplitudes
\begin{equation}
  \bm z(t)=e^{-iMt}\bm z(0).
  \label{eq:app-prop-ampl}
\end{equation}
Equivalently, for the normally ordered number-conserving Hamiltonian of Eq.~\eqref{eq:H}, which contains no zero-point constant in the generator of the dynamics,
\[
  e^{-i\hat Ht/\hbar}\ket{\bm z(0)}
  =\ket{\bm z(t)}.
\]
If a state-independent zero-point constant were restored in the total Hamiltonian, it would contribute only a common global phase to both branches and would therefore leave all reduced density matrices, overlaps, ergotropies, and concurrences unchanged. Applying Eq.~\eqref{eq:app-prop-ampl} separately to the two initial branch vectors $(\alpha_0,0,0,0)^T$ and $(\beta_0,0,0,0)^T$ gives the amplitudes $\alpha_t,\lambda_t^{(i)}$ and $\beta_t,\chi_t^{(i)}$ used in Eq.~\eqref{eq:psit}. Linearity of Eq.~\eqref{eq:app-prop-ampl} is what preserves the two-branch coherent-state structure throughout the dynamics.

In the symmetric resonant case, the single-particle dynamics separates into one bright battery mode and two dark modes. Since the battery is initially in vacuum, the dark modes remain unpopulated and the amplitudes satisfy
\begin{equation}
  i\dot f=3\gamma g,\qquad i\dot g=\gamma f,
  \qquad f(0)=1,\quad g(0)=0.
\end{equation}
The solution is
\[
  f(t)=\cos(\sqrt3\,\gamma t),\qquad
  g(t)=-\frac{i}{\sqrt3}\sin(\sqrt3\,\gamma t),
\]
which is Eq.~\eqref{eq:fg}. The same argument for $N$ identical resonant cells gives a bright-mode coupling $\sqrt N\gamma$, hence
\[
  f(t)=\cos(\sqrt N\,\gamma t),\qquad
  g(t)=-\frac{i}{\sqrt N}\sin(\sqrt N\,\gamma t).
\]
For non-symmetric configurations, Eq.~\eqref{eq:app-prop-ampl} remains exact. The matrix exponential therefore supplies all amplitudes required in the main text, even when no closed trigonometric form is available.

\section{Energy expectation values}\label{app:energy}
This appendix gives the stored-energy expressions used in Eqs.~\eqref{eq:WB} and~\eqref{eq:WBi}. Since the zero-point contribution has been separated as $E_0\mathbb I_B$, only
\[
  \widetilde{\hat H}_B=\hbar\sum_{i=1}^3\omega_i\hat b_i^\dagger\hat b_i
\]
enters $\Delta E_B$. For coherent product states,
\begin{equation}
  \langle\chi_t|\hat b_i^\dagger\hat b_i|\lambda_t\rangle
  =\chi_t^{(i)*}\lambda_t^{(i)}q_B^*.
  \label{eq:app-cross-global}
\end{equation}
Substitution of Eq.~\eqref{eq:rhoB} gives
\begin{align}
  \Delta E_B&=\hbar\varepsilon_B,\\
  \varepsilon_B&=|\mathcal N_0|^2\Bigl[|c_\alpha|^2\sum_i\omega_i|\lambda_t^{(i)}|^2
  +|c_\beta|^2\sum_i\omega_i|\chi_t^{(i)}|^2\nonumber\\
  &\hspace{1.0cm}+2\operatorname{Re}\!\left(c_\alpha c_\beta^*s^*q_B^*
  \sum_i\omega_i\chi_t^{(i)*}\lambda_t^{(i)}\right)\Bigr].
  \label{eq:app-XiB}
\end{align}

For one cell,
\begin{equation}
 \Delta E_{B_i}=\hbar\omega_i\bar n_i.
\end{equation}
The interference matrix element on the retained cell is
\begin{equation}
 \langle\chi_t^{(i)}|\hat b_i^\dagger\hat b_i|\lambda_t^{(i)}\rangle
 =\chi_t^{(i)*}\lambda_t^{(i)}q_i^*.
 \label{eq:local-cross-correct}
\end{equation}
Combining this factor with the overlap $Q_i^*$ generated by tracing out the other two cells yields
\begin{align}
  \bar n_i=|\mathcal N_0|^2\Bigl[&|c_\alpha|^2|\lambda_t^{(i)}|^2
  +|c_\beta|^2|\chi_t^{(i)}|^2\nonumber\\
  &+2\operatorname{Re}\!\left(c_\alpha c_\beta^*s^*Q_i^*q_i^*
      \chi_t^{(i)*}\lambda_t^{(i)}\right)\Bigr].
  \label{eq:app-nibar}
\end{align}
Since $Q_iq_i=q_B$, summing $\hbar\omega_i\bar n_i$ over the cells reproduces Eq.~\eqref{eq:app-XiB} exactly, as required by additivity of the excitation Hamiltonian.

In the symmetric resonant even-cat protocol, the expression reduces to
\begin{equation}
 \Delta E_B=\hbar\omega\bar n_0\Theta(t),\qquad
 \Delta E_{B_i}=\frac13\Delta E_B,
\end{equation}
with $\bar n_0=a\tanh a$.

\section{Local coherent and incoherent ergotropy of a single cell}\label{app:localcohinc}
We now prove the passivity statement used in Sec.~\ref{sec:cohinc}. Projecting Eq.~\eqref{eq:rhoBi} onto the Fock basis gives
\begin{align}
  p_n^{(i)}=|\mathcal N_0|^2\Bigl[&|c_\alpha|^2p_{\lambda_i}(n)+|c_\beta|^2p_{\chi_i}(n)\nonumber\\
  &+2\operatorname{Re}\!\left(c_\alpha c_\beta^*s^*Q_i^*K_i(n)\right)\Bigr],
\end{align}
where $p_z(n)=e^{-|z|^2}|z|^{2n}/n!$ and
\begin{equation}
 K_i(n)=e^{-(|\lambda_i|^2+|\chi_i|^2)/2}
 \frac{(\lambda_i\chi_i^*)^n}{n!}.
\end{equation}
For a single oscillator the energy spectrum is nondegenerate, so the dephased state is passive exactly when the sequence $p_n^{(i)}$ is nonincreasing.

For the symmetric resonant even cat, let
\begin{equation}
 m\equiv m_i=\frac{ay}{3},\qquad
 r\equiv r_i=e^{-2a(1-y/3)}=e^{-2(a-m)},
\end{equation}
where $y=\Theta\in[0,1]$. Then
\begin{equation}
 p_n^{(i)}=\frac{e^{-m}m^n}{n!(1+e^{-2a})}\,[1+(-1)^nr].
 \label{eq:localcatpops-app}
\end{equation}
The ratio of consecutive populations is
\begin{equation}
  \frac{p_{n+1}^{(i)}}{p_n^{(i)}}=\frac{m}{n+1}
  \begin{cases}
    \tanh(a-m),& n\ \text{even},\\[4pt]
    \coth(a-m),& n\ \text{odd}.
  \end{cases}
  \label{eq:localratio}
\end{equation}
For fixed $m$, the largest even-$n$ ratio occurs at $n=0$ and the largest odd-$n$ ratio at $n=1$. Define
\begin{equation}
 F_e(m)=m\tanh(a-m),\qquad
 F_o(m)=\frac m2\coth(a-m).
\end{equation}
On $0\le m\le a/3$, both functions are increasing. Indeed,
\begin{equation}
 F_e'(m)=\frac{\sinh(a-m)\cosh(a-m)-m}{\cosh^2(a-m)}>0,
\end{equation}
because $z=a-m\ge2m\ge m$ and $\sinh z\cosh z>z$ for $z>0$, while
\begin{equation}
 F_o'(m)=\frac12\left[\coth(a-m)+m\operatorname{csch}^2(a-m)\right]>0.
\end{equation}
Therefore the worst case occurs at $m=a/3$ ($y=1$), and the diagonal state is passive throughout the charging branch if and only if
\begin{equation}
 \frac a3\tanh\!\left(\frac{2a}{3}\right)\le1,
 \qquad
 \frac a6\coth\!\left(\frac{2a}{3}\right)\le1.
 \label{eq:passivityexact}
\end{equation}
The first inequality becomes limiting first, at
\begin{equation}
 a_c\simeq3.098007208.
\end{equation}
For the value used in the figures, $a=2.25$,
\begin{equation}
 F_e(a/3)\simeq0.67886,\qquad
 F_o(a/3)\simeq0.41430,
\end{equation}
so $p_{n+1}^{(i)}<p_n^{(i)}$ for every $n$ and every $0<y\le1$. Hence
\begin{equation}
 \Wb_{B_i}^{\rm inc}=0,\qquad
 \Wb_{B_i}^{\rm coh}=\Wb_{B_i}.
\end{equation}
The criterion is controlled by the combined cat parameters through Eq.~\eqref{eq:passivityexact}; it is not equivalent to the simpler condition $m<1$. Under uniform loss, the same proof applies with $y\to\widetilde\Theta\in[0,1]$.

\section{Effective two-qubit reduction, determinant, and Wootters concurrence}\label{app:tangles}
The oscillator Hilbert spaces are infinite dimensional, but every relevant party occupies the span of two normalized branch states, a structure familiar from entangled coherent states~\cite{Sanders1992,Sanders2012}. Let $\ket{a_X}$ and $\ket{b_X}$ denote the two branch states of a party $X$ and let
\begin{equation}
 q_X=\langle a_X|b_X\rangle.
\end{equation}
For $|q_X|<1$, Gram--Schmidt gives the exact orthonormal basis
\begin{equation}
 \ket{0_X}=\ket{a_X},\qquad
 \ket{1_X}=\frac{\ket{b_X}-q_X\ket{a_X}}{\sqrt{1-|q_X|^2}},
 \label{eq:GSbasis}
\end{equation}
so that
\begin{equation}
 \ket{b_X}=q_X\ket{0_X}+\sqrt{1-|q_X|^2}\ket{1_X}.
\end{equation}
The case $q_X=0$ is completely regular: the two branch states are already orthogonal. A true reduction of the support dimension occurs only at $|q_X|=1$, when the branch states are linearly dependent; all formulas below then follow by continuity and the corresponding concurrence involving $X$ vanishes.

For completeness, consider a normalized rank-two operator
\begin{equation}
 \rho=\mathcal Z^{-1}\left(A\ket u\bra u+B\ket v\bra v+D\ket u\bra v+D^*\ket v\bra u\right),
 \label{eq:genericrank2}
\end{equation}
where $r=\langle u|v\rangle$. Let $U=(\ket u,\ket v)$,
\begin{equation}
 K=\begin{pmatrix}A&D\\D^*&B\end{pmatrix},\qquad
 G=U^\dagger U=\begin{pmatrix}1&r\\r^*&1\end{pmatrix}.
\end{equation}
On the support of $\rho$,
\begin{equation}
 \det\rho=\mathcal Z^{-2}\det K\det G
 =\mathcal Z^{-2}(AB-|D|^2)(1-|r|^2).
 \label{eq:genericdet}
\end{equation}
Applying this identity to $\rho_B$ gives Eq.~\eqref{eq:detB}; applying it to $\rho_{B_i}$ gives Eq.~\eqref{eq:detBi}.

We next derive the concurrence formula in a way that covers both unitary and dissipative dynamics. Consider a bipartition $A|B$ after a remainder $R$ has been traced out. In the orthonormal bases~\eqref{eq:GSbasis}, the two product branch vectors are $\ket u=\ket{a_Aa_B}$ and $\ket v=\ket{b_Ab_B}$, and the reduced state has the form
\begin{equation}
 \rho_{AB}=p\ket u\bra u+q\ket v\bra v+z\ket u\bra v+z^*\ket v\bra u,
 \label{eq:rhoABgeneric}
\end{equation}
with
\begin{equation}
 p=|\mathcal N_0|^2|c_\alpha|^2,\qquad
 q=|\mathcal N_0|^2|c_\beta|^2,
\end{equation}
and
\begin{equation}
 z=|\mathcal N_0|^2c_\alpha c_\beta^*w\,q_R^*.
 \label{eq:zgeneric}
\end{equation}
Here $w=1$ in unitary evolution and $w=e^{-\Gamma}$ in the dissipative problem. In the effective two-qubit representation, the spin flip satisfies
\begin{equation}
 \langle u|\widetilde u\rangle=\langle v|\widetilde v\rangle=0,
\end{equation}
and
\begin{equation}
 \left|\langle u|\widetilde v\rangle\right|
 =\sqrt{1-|q_A|^2}\sqrt{1-|q_B|^2}
 \equiv\kappa_{AB}.
\end{equation}
A direct evaluation of $\rho_{AB}\widetilde\rho_{AB}$ gives two nonzero eigenvalues,
\begin{equation}
 \lambda_\pm=\kappa_{AB}^2\left(\sqrt{pq}\pm|z|\right)^2,
 \label{eq:lambdapm}
\end{equation}
with the other two equal to zero. Positivity implies $|z|\le\sqrt{pq}$, so Wootters' formula yields
\begin{align}
 \Cc_{A:B}
 &=\sqrt{\lambda_+}-\sqrt{\lambda_-}\nonumber\\
 &=2|z|\kappa_{AB}\nonumber\\
 &=2|\mathcal N_0|^2|c_\alpha c_\beta|\,|w|\,|q_R|
 \sqrt{1-|q_A|^2}\sqrt{1-|q_B|^2}.
 \label{eq:Cmasterappendix}
\end{align}
Thus every concurrence in the manuscript is a Wootters concurrence on an exact $2\times2$ support. In the unitary case $|w|=1$, Eq.~\eqref{eq:Cmasterappendix} reduces to Eq.~\eqref{eq:Cgeneral}; with dissipation, $|w|^2=x_E$ and its square gives Eq.~\eqref{eq:CgeneralLoss}. For a pure unitary cut with $R=\varnothing$, Eq.~\eqref{eq:Cmasterappendix} also reduces to $\Cc=2\sqrt{\det\rho_A}=2\sqrt{\det\rho_B}$.

\section{Closure under linear dissipation}
\label{app:diss}

We verify Eqs.~\eqref{eq:rhodiss} and~\eqref{eq:Gamma}. From the same pure initial state \eqref{eq:init}, written as 
\begin{equation*}
    \rho (0) = |\mathcal N_0|^2 \sum_{A,B\in\{\alpha,\beta\}} c_Ac_B^*\ket{\bm z^A(0)}\bra{\bm z^B(0)},
\end{equation*}
where the branch initial amplitudes are
\begin{equation}
 \bm z^\alpha(0)=(\alpha_0,0,0,0)^T,\qquad
 \bm z^\beta(0)=(\beta_0,0,0,0)^T.
 \label{eq:diss-branch-initial}
\end{equation}
We use the normalized coherent dyad $\sigma_{AB}(t)=\ket{\bm z^A(t)}\bra{\bm z^B(t)}$ and assume the ansatz
\begin{equation}
    \rho (t) = |\mathcal N_0|^2\sum_{A,B\in\{\alpha,\beta\}} c_A c_B^* w_{AB}(t) \sigma_{AB}(t),
\end{equation}
on which $w_{AB}(t)$ represent the non-diagonal dissipative coefficients needed to solve the Lindblad equation \eqref{eq:lindblad}:
\begin{equation*}
  \dot\rho = -\frac{i}{\hbar}[\hat H,\rho]
  + \sum_{\mu\in\{c,1,2,3\}} \kappa_\mu
    \Bigl( \hat a_\mu \rho\, \hat a_\mu^\dagger
    - \tfrac12 \{ \hat a_\mu^\dagger \hat a_\mu , \rho \} \Bigr).
\end{equation*}

Substituting the coherent dyad $\sigma_{AB}$ into the Lindblad equation produces the linearly independent operator structures $\{\hat a_\mu^\dagger\sigma_{AB},\sigma_{AB}\hat a_\nu,\sigma_{AB}\}$ on both sides. Equating their coefficients gives the amplitude equations
\begin{equation}
    \dot{\bm z}^{A,B}(t) = -(iM + \Lambda)\bm z^{A,B}(t),
\end{equation}
and
\begin{equation}
    \frac{\dot w_{AB}}{w_{AB}} = -\frac12 \sum_\mu \kappa_\mu
  \Bigl[|z^A_\mu|^2+|z^B_\mu|^2 - 2z^A_\mu z^{B*}_\mu\Bigr]
  \label{eq:wode}
\end{equation}
for the branch weights. For $A=B$ this equation preserves $w_{AA}=1$; for $A\neq B$ its integration gives the decoherence exponent $\Gamma(t)$ used in the main text. Here, $\Lambda=\tfrac12\mathrm{diag}(\kappa_c,\kappa_1,\kappa_2,\kappa_3)$, and $M$ is the bilinear coupling of the Hamiltonian $\hat H=\hbar\sum_{\mu,\nu}M_{\mu,\nu}\hat a_\mu^\dagger \hat a_\nu$ -- the same matrix that controls the unitary evolution \eqref{eq:M}. The dissipative matrix $\Lambda$ modifies the branch amplitudes but does not enlarge the two-branch support, which is why the rank-two structure survives.

In the symmetric resonant case with uniform loss, with $\gamma_i=\gamma$, $\kappa_i=\kappa$, and $\omega_i=\omega$, the solution is
\begin{align*}
    z^A_c(t)&=\alpha_0 e^{-\left(i\omega + \kappa/2\right)t}\cos(\sqrt{3}\gamma t),\\
    z^A_\mu(t)&=\frac{-i\alpha_0}{\sqrt{3}} e^{-\left(i\omega + \kappa/2\right)t}\sin(\sqrt{3}\gamma t)\quad\text{for $\mu=1,2,3$},
\end{align*}
and similarly for the $\beta$ branch.
\begin{align}
    \Gamma(t) = \tfrac12 \left(|\alpha_0|^2 + |\beta_0|^2 - 2\alpha_0\beta_0^*)(1-e^{-\kappa t}\right)
\end{align}
Specializing to the even-cat initial condition $\alpha_0=-\beta_0$ gives the expressions used in the main text.

Trace preservation provides a useful consistency check. For the off-diagonal branch,
\begin{equation}
 w_{\alpha\beta}(t)\langle\bm z^\beta(t)|\bm z^\alpha(t)\rangle
 =\langle\bm z^\beta(0)|\bm z^\alpha(0)\rangle,
\end{equation}
so the constant normalization $\mathcal N_0$ of the initial cat remains the normalization entering Eq.~\eqref{eq:rhodiss}; environmental decoherence is carried by $w_{\alpha\beta}$ rather than by a time-dependent overall prefactor.

\section{Monotonicity of global and local ergotropies}\label{app:mono}
We prove the monotonicity statements used in Sec.~\ref{sec:symmetric}. Let $y=\Theta\in[0,1]$ for the symmetric resonant even-cat charging branch.

\paragraph{Global ergotropy.}
From Eqs.~\eqref{eq:WB} and~\eqref{eq:detB},
\begin{equation}
 \Wb_B=\hbar\omega(\bar n_0y-\nu_-),
\end{equation}
with
\begin{equation}
 \det\rho_B=\frac{\mathcal A}{4}(1-x_C)(1-x_B),
\end{equation}
where $x_C=e^{-4a(1-y)}$, $x_B=e^{-4ay}$, $\mathcal A=(1+e^{-2a})^{-2}$, and $\bar n_0=a\tanh a$. Since $x_Cx_B=e^{-4a}$,
\begin{equation}
 \frac{d}{dy}\det\rho_B=a\mathcal A(x_B-x_C).
\end{equation}
Using $d\nu_-/dy=(d\det\rho_B/dy)/\sqrt{1-4\det\rho_B}$ gives
\begin{equation}
 \frac{1}{\hbar\omega}\frac{d\Wb_B}{dy}
 =\bar n_0-\frac{a\mathcal A(x_B-x_C)}{\sqrt{1-4\det\rho_B}}.
 \label{eq:globalmonoderiv}
\end{equation}
Set $s_0=e^{-2a}$ and $x_C=s_0e^{-t}$, $x_B=s_0e^t$, with $t=2a(1-2y)$. For $t<0$ the second term in Eq.~\eqref{eq:globalmonoderiv} is negative and the derivative is positive. For $t\ge0$, squaring the equivalent nonnegative inequality reduces it to
\begin{equation}
 (1-s_0)^2\ge2s_0(\cosh t-1).
\end{equation}
The right-hand side is maximal at $t=2a$ ($y=0$), where equality holds. Hence
\begin{equation}
 \frac{d\Wb_B}{dy}\ge0
\end{equation}
for $0\le y\le1$, with equality only at $y=0$.

\paragraph{Single-cell ergotropy.}
For one cell,
\begin{equation}
 \Wb_{B_i}=\hbar\omega\left(\frac{\bar n_0}{3}y-\nu_-^{(i)}\right).
\end{equation}
Define
\begin{equation}
 x=e^{-4ay/3},\qquad z=e^{-4a(1-y/3)}.
\end{equation}
Then $xz=e^{-4a}$ and
\begin{equation}
 D_i\equiv\det\rho_{B_i}=\frac{\mathcal A}{4}(1-x)(1-z),
\end{equation}
so
\begin{equation}
 D_i'=\frac{a\mathcal A}{3}(x-z).
\end{equation}
Let $s_0=e^{-2a}$ and
\begin{equation}
 t=2a\left(1-\frac{2y}{3}\right),\qquad
 x=s_0e^t,\quad z=s_0e^{-t}.
\end{equation}
A direct simplification gives
\begin{equation}
 \sqrt{1-4D_i}=\frac{\cosh(t/2)}{\cosh a}
\end{equation}
and therefore
\begin{equation}
 \frac{d\nu_-^{(i)}}{dy}
 =\frac{a}{3\cosh a}\sinh\!\left[a\left(1-\frac{2y}{3}\right)\right].
\end{equation}
Since $\bar n_0=a\tanh a$, we obtain
\begin{equation}
 \frac{1}{\hbar\omega}\frac{d\Wb_{B_i}}{dy}
 =\frac{a}{3\cosh a}
 \left[\sinh a-\sinh\!\left(a\left(1-\frac{2y}{3}\right)\right)\right].
 \label{eq:localmonoproof}
\end{equation}
For $a>0$ and $0<y\le1$, the argument of the second hyperbolic sine lies strictly below $a$, so Eq.~\eqref{eq:localmonoproof} is strictly positive. The derivative vanishes only at $y=0$. Therefore both global and local ergotropies reach their first charging-branch maxima at $y=1$, i.e. at $t_*$. Under uniform loss, the same battery-only formulas hold with $y\to\widetilde\Theta$, so their maxima occur at the maximum of $\widetilde\Theta$, namely $t_{\rm opt}$.

\end{document}